\documentclass[%
 reprint,
 amsmath,amssymb,
 aps,
 prf,
 onecolumn,
 superscriptaddress,
]{revtex4-2}

\usepackage{graphicx}
\usepackage{dcolumn}
\usepackage{bm}

\usepackage{xcolor}
\usepackage{float}

\begin{document}


\title{Coalescence of viscoelastic sessile drops: the small and large contact angle limits }

Published in JFM: https://doi.org/10.1017/jfm.2025.10878

 \author{Paul R. Kaneelil}
    \affiliation{Department of Mechanical and Aerospace Engineering, Princeton University, Princeton, New Jersey 08544, USA}
 \author{Kazuki Tojo}
    \affiliation{Department of Mechanical and Aerospace Engineering, Princeton University, Princeton, New Jersey 08544, USA}
 \author{Palas Kumar Farsoiya}
    \affiliation{Department of Mechanical and Aerospace Engineering, Princeton University, Princeton, New Jersey 08544, USA}
    
 \author{Luc Deike}
    \affiliation{Department of Mechanical and Aerospace Engineering, Princeton University, Princeton, New Jersey 08544, USA}
    \affiliation{High Meadows Environmental Institute, Princeton University, Princeton, NJ 08544, USA}
 \author{Howard A. Stone}
    \affiliation{Department of Mechanical and Aerospace Engineering, Princeton University, Princeton, New Jersey 08544, USA}

\date{January 10, 2026}

\begin{abstract}
The coalescence and breakup of drops are classic examples of flows that feature singularities. The behavior of viscoelastic fluids near these singularities is particularly intriguing—not only because of their added complexity, but also due to the unexpected responses they often exhibit. In particular, experiments have shown that the coalescence of viscoelastic sessile drops can differ significantly from their Newtonian counterparts, sometimes resulting in a sharply defined interface. However, the mechanisms driving these differences in dynamics, as well as the potential influence of the contact angle are not fully known. Here, we study two different flow regimes effectively induced by varying the contact angle and demonstrate how that leads to markedly different coalescence behaviors. We show that the coalescence dynamics is effectively unaltered by viscoelasticity at small contact angles. The Deborah number, which is the ratio of the relaxation time of the polymer to the timescale of the background flow, scales as $\theta^3$ for $\theta \ll 1$, thus rationalizing the near-Newtonian response. On the other hand, it has been shown previously that viscoelasticity dramatically alters the shape of the interface during coalescence at large contact angles. We study this large contact angle limit using experiments and 2D numerical simulations of the equation of motion. We show that the departure of the coalescence dynamics from the Newtonian case is a function of the Deborah number and the elastocapillary number, which is the ratio between the shear modulus of the polymer solution and the characteristic stress in the fluid.

\end{abstract}

\maketitle

\section{Introduction} \label{Ch6sec:intro}

Viscoelastic fluids are a class of non-Newtonian fluids that exhibit a viscous and an elastic response to the application of stress \citep{snoeijer2020relationship}. Some examples include biological fluids such as saliva \citep{mitchinson2010saliva}, the childrens' toy Silly Putty, and other complex fluids such as surfactant solutions and emulsions. The common feature that is present in these different viscoelastic fluids is that they all contain molecules that are able to stretch and relax over some time scale. Polymer solutions are a common viscoelastic fluid and have been used extensively to study the influence of material properties on flow behavior and the many interesting phenomena that arise as they flow \citep{datta2022perspectives}. For example, the addition of polymers to a solution leads to drag reduction in turbulent pipe flows \citep{Toms1948SomeOO,berman1978drag} and can alter drop breakup by producing long and stable threads \citep{anna2001elasto}.

\begin{figure}
\begin{center}
\includegraphics[width=0.8\textwidth]{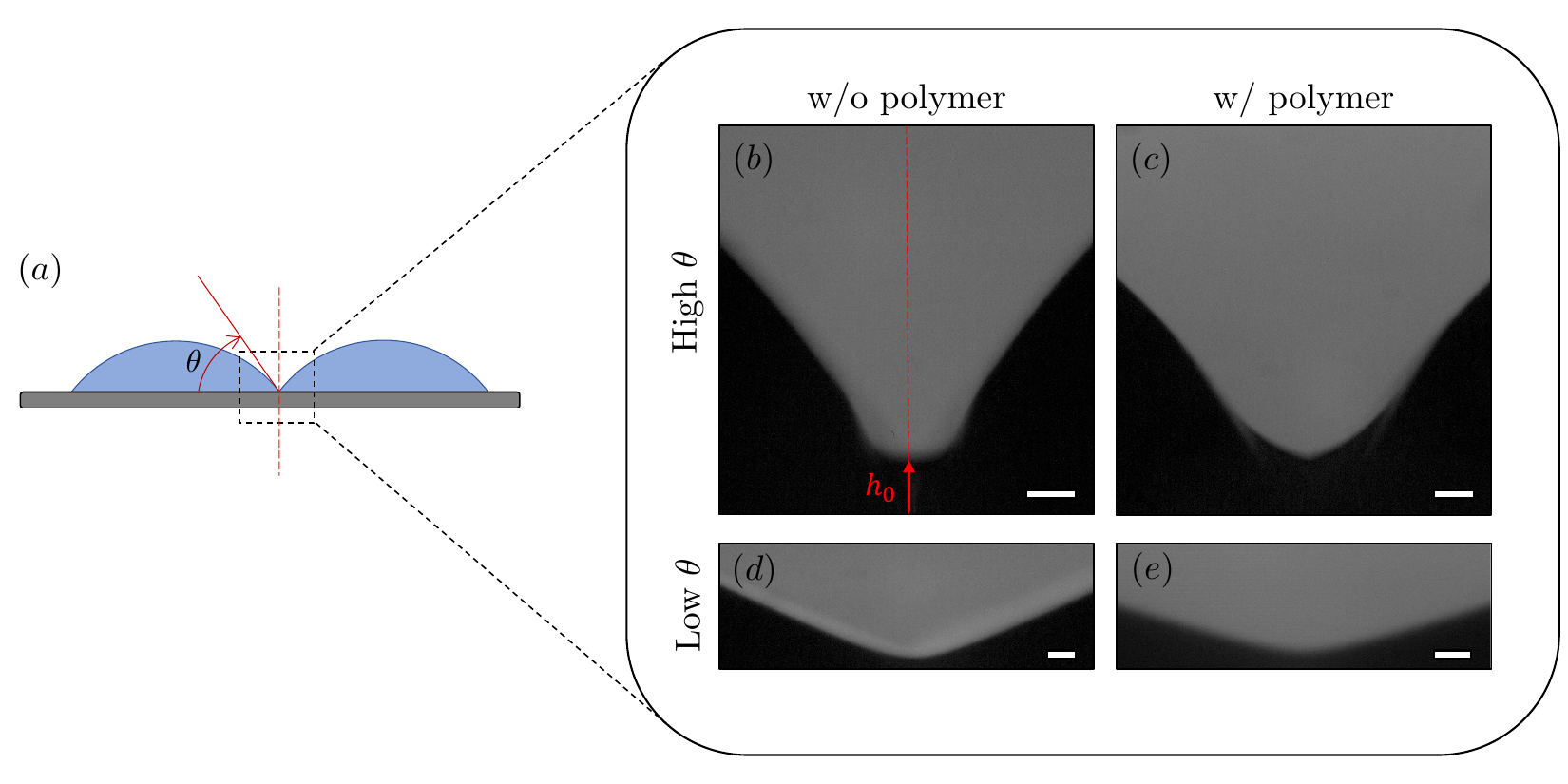}
\caption{\label{ch6fig1} The shape of the interface during the coalescence of Newtonian and polymeric drops. (a) Schematic of the side view profile of two drops during a typical coalescence experiment. Experimental images of large $\theta$ coalescence of (b) water and (c) 0.5\% PEO drops show a significant difference in the shape of the interface. On the other hand, small $\theta$ coalescence of (d) 1000 cSt silicone oil and (e) 0.5\% PEO drops show similar shape of the interface. Scale bars represent 0.1 mm. }
\end{center}
\end{figure}

More recently, the coalescence of drops of polymer solution has been studied under different configurations varying both the drop geometry and the substrate wettability \citep{varma2020universality,varma2021coalescence,dekker2022elasticity,chen2022probing,varma2022elasticity,varma2022rheocoalescence,sivasankar2023coalescence,eggers2024reviewcoalescence,rostami2025coalescence}. Here, we will focus on the coalescence of sessile drops (Figure~\ref{ch6fig1}). Studies have shown that in the case of semi-dilute polymer solutions and inertially dominated coalescence, which often correspond to a large contact angle $\theta$, the temporal evolution of the bridge height $h_0$ (see Fig. \ref{ch6fig1}) follows the same scaling as its Newtonian counterpart: $h_0 \propto t^{2/3}$ \citep{varma2022elasticity,dekker2022elasticity}. While the temporal scaling of the bridge height was unaffected, the shape of the interface was highly altered by the polymer stress. The Newtonian, self-similar scaling that describes the evolution of the interface at early times after coalescence breaks down for these semi-dilute polymer solutions and different self-similar scalings were proposed taking into account the stress induced by the polymers \citep{dekker2022elasticity, varma2022elasticity}. As the polymer concentration increased, the magnitude of the temporal scaling exponent decreased from $\frac{2}{3}$ to about $0.5$ in certain cases and even lower in others \citep{varma2021coalescence,varma2022elasticity,chen2022probing,rostami2025coalescence}. The nature of the change in the scaling exponent was shown to be affected by the way in which the drops were deposited. When two drops of fixed volume are deposited by allowing them to impact the substrate and spread until they come into contact, the scaling exponent decreases to approximately $0.5$ with increasing polymer concentration. When fluid is injected into two drops so that their volume increases until they touch and coalesce, the scaling exponent becomes even lower. \citep{varma2022elasticity}. While much is known about the effect of elasticity on the coalescence dynamics, a comprehensive understanding of the departure of these dynamics from the Newtonian dynamics is still lacking. To the best of our knowledge, the coalescence of viscoelastic sessile drops in the small contact angle regime also remains unexplored.

Here, we combine laboratory experiments and numerical simulations using the Oldroyd-B model to investigate the coalescence of sessile viscoelastic drops at small and large static advancing contact angle of the drops at the moment of coalescence, hereafter referred to simply as the contact angle $\theta$. Experimentally, we consider the coalescence of sessile semi-dilute polymeric drops (polyethylene glycol PEO; molecular weight $M_w=4 \times 10^6$ g/mol). To illustrate the influence of the contact angle, Figure \ref{ch6fig1} shows side view profiles of the coalescence of drops with and without polymer at large and small contact angles. The difference in the shape of the interface with and without polymer is clearly evident at large contact angles, as shown in Fig. \ref{ch6fig1}(b)-(c), and as previously reported \citep{dekker2022elasticity}. In contrast, the shape of the interface with and without polymer at small contact angles looks nearly the same, as shown in Fig. \ref{ch6fig1}(d)-(e). 

In what follows, in Section \ref{seclowangle}, we analytically and experimentally show that the coalescence dynamics is virtually unaffected by the presence of polymers in the limit $\theta \ll 1$, which we show is connected to the Deborah number, $De$, being small. The Deborah number is the ratio between the relaxation time of the polymer and the characteristic time scale of the background flow. In section \ref{sechighangle}, we use numerical simulations to study coalescence at $\theta \gtrapprox 1$. We show that the Oldroyd-B model can capture the effect of polymers seen in the experiments and further illustrate the coalescence dynamics as a function of the Deborah number and the elastocapillary number, $Ec$, of the polymer solution. The elastocapillary number is the ratio between the shear modulus of the polymer solution and the characteristic stress in the fluid. Exploring the various limits of the Oldroyd-B model, we explain our numerical results and provide more insight on the departure of the coalescence dynamics from the Newtonian case. In both our experiments and simulations, we make sure that the timescale of coalescence is clearly separated from all other relevant timescales. For instance, in the experiments, the droplets spread slowly over several minutes prior to contact, while the actual coalescence occurs over just a few seconds. In simulations, this feature is straightforward to achieve by initializing all velocities to zero.
 
\section{Governing equations} 
\label{goveq}
Here, we introduce the general equations that will be rescaled later to identify the dimensionless parameters for the coalescence problem. The governing mass and momentum equations for incompressible flows are given by
\begin{subequations}
\label{Eqmassmomen}
\begin{gather}
\Tilde{\nabla} \cdot \Tilde{\pmb{u}}=0, \label{eqch6continuity}
\\
\rho \frac{\textrm{D} \Tilde{\pmb{u}}}{\textrm{D} \Tilde{t}} = -\Tilde{\nabla} \Tilde{p} + \Tilde{\nabla} \cdot \Tilde{\pmb{\tau}}, \label{eqch6momentum}
\end{gather}
\end{subequations}where $\frac{\textrm{D} }{\textrm{D} \Tilde{t}}$ denotes the material derivative; the tilde is used to represent dimensional variables. Here, $\Tilde{\pmb{u}}$ is the velocity field, $\Tilde{p}$ is the pressure, $\rho$ is the density, and $\Tilde{\pmb{\tau}}$ is the deviatoric stress tensor. For polymeric fluids, the deviatoric stress tensor can be written as the sum of the solvent and polymeric contributions: $\Tilde{\pmb{\tau}} = \Tilde{\pmb{\tau}}^s + \Tilde{\pmb{\tau}}^{p}$. The solvent contribution is the Newtonian stress $\Tilde{\pmb{\tau}}^s = 2\mu_s \Tilde{\pmb{E}}$, where $\mu_s$ is the solvent viscosity and $\Tilde{\pmb{E}} = \frac{1}{2}[(\Tilde{\nabla} \Tilde{\pmb{u}})+(\Tilde{\nabla} \Tilde{\pmb{u}})^T]$ is the rate of strain tensor. The polymer contribution will be modelled by the Oldroyd-B model according to

\begin{subequations}
    \label{oldroydBgeneral}
    \begin{gather}
        \Tilde{\pmb{\tau}}^{p} = G (\pmb{A} - \pmb{I}), \label{oldroydBgeneral1}
        \\
        \overset{\mathtt{\nabla}}{\pmb{A}} = -\frac{1}{\lambda}(\pmb{A} - \pmb{I}), \label{oldroydBgeneral2}
        \\
        \lambda \overset{\mathtt{\nabla}}{\Tilde{\pmb{\tau}}^p} + \Tilde{\pmb{\tau}}^p = 2 G \lambda \Tilde{\pmb{E}},\label{oldroydBgeneral3}
    \end{gather}
\end{subequations}where, $G$ is the elastic modulus of the polymer solution, $\pmb{A}$ is the conformation tensor of the polymer, $\pmb{I}$ is the identity tensor, and $\lambda$ is the relaxation time of the polymer solution. The upper convected derivative is defined as $\overset{\mathtt{\nabla}}{\pmb{A}} = \frac{D}{D \tilde{t}}\pmb{A}  - (\Tilde{\nabla} \Tilde{\pmb{u}})^T \cdot \pmb{A} - \pmb{A} \cdot \Tilde{\nabla} \Tilde{\pmb{u}}$. Note that Eq. (\ref{oldroydBgeneral1}) is a state of stress assumption and Eq. (\ref{oldroydBgeneral2}) is an evolution equation for the microstructure of the polymer chains, which are allowed to stretch/compress under flow and relax with the longest relaxation time $\lambda$ \citep{stone2023note}. Equation (\ref{oldroydBgeneral3}) is usually referred to as the Oldroyd-B model, and can be obtained by taking the upper convected derivative of Eq. (\ref{oldroydBgeneral1}) and appropriately substituting Eq. (\ref{oldroydBgeneral2}) \citep{snoeijer2020relationship}.

\section{The small contact angle limit}
\label{seclowangle}

We start by considering the coalescence of polymeric drops in the small contact angle or the thin-film limit. We identify the important dimensionless parameters governing the dynamics and show experimental results from imaging the three-dimensional shape of the interface during coalescence.

\subsection{Dimensionless equations and parameters}
\label{seclowangletheory}
In the small contact angle limit, $\theta \ll 1$, the lubrication approximation has been shown to successfully describe the dynamics of coalescing sessile Newtonian drops \citep{ristenpart2006coalescence,hernandez2012symmetric,kaneelil2022three}. Therefore, the equations can be rescaled using the following variables $x=\Tilde{x}/\ell_c,~y=\Tilde{y}/\ell_c,~z=\Tilde{z}/(\theta\ell_c),~u_x=\Tilde{u_x}/u_c,~u_y=\Tilde{u_y}/u_c,~u_z=\Tilde{u_z}/(\theta u_c),~t=\Tilde{t}/(\ell_c/u_c)$ and $p=\Tilde{p}\theta^2 \ell_c/(\mu u_c)$, where $x$ and $y$ are the in-plane coordinates on the substrate and $z$ is the out-of-plane coordinate, as shown in Fig. \ref{ch6fig2}(a). Also, $\ell_c=\sqrt{\gamma/(\rho g)}$ is the capillary length, which is a good approximation for the largest height of the drop in our experiments, $\mu$ is the viscosity of the solution, and $u_c=\gamma \theta^3/(3\mu)$ is the velocity scale for coalescence \citep{hernandez2012symmetric}. Also, $Oh_{\theta}=\mu/\sqrt{\rho \gamma \ell_c}$ denotes the Ohnesorge number. Note that we neglect inertial effects in this limit, as usual, since the inertial terms in the momentum equation Eq. (\ref{eqch6momentum}) will vary, according to the above scalings, as $\theta^5/Oh_{\theta}^2$ (about $10^{-6}$ in our experiments) and will always be small when $\theta \ll 1$. 

Rescaling the Oldroyd-B equations with these variables gives
\begin{subequations}
\label{low_oldroydB}
\begin{gather}
\pmb{\tau}^p=3 Ec_{\theta}(\pmb{A}-\pmb{I}), \label{low_oldroydB1}
\\
\overset{\mathtt{\nabla}}{\pmb{A}}=-\frac{1}{De_\theta}(\pmb{A}-\pmb{I}), \label{low_oldroydB2}
\end{gather}
\end{subequations} where the Deborah number and the elastocapillary number are defined as
\begin{equation}
    De_{\theta} = \frac{\lambda \gamma \theta^3}{3\mu\ell_c}, ~~~\textrm{and}~~~ Ec_{\theta}=\frac{G\ell_c}{\theta \gamma},
    \label{eq_smalltheta_de_ec}
\end{equation}
with the subscript $\theta$ indicating the $\theta \ll 1$ limit. Similarly, we rescale Eq. (\ref{oldroydBgeneral3}) and focus on the $z$-components of the equations, which become
\begin{subequations}
\begin{gather}
De_\theta~\overset{\mathtt{\nabla}}{\tau^p_{zj}} +  \tau^p_{zj} = 3~ \frac{Ec_{\theta} De_\theta}{\theta}~ \frac{\partial u_j}{\partial z} + {\cal O}(\theta^2),~~~~j = x,y  \label{low_oldroydB3a}
\\
De_\theta~\overset{\mathtt{\nabla}}{\tau^p_{zj}} +  \tau^p_{zj} = 6~ Ec_{\theta} De_\theta~ \frac{\partial u_j}{\partial z},~~~~~~~~~~~~~~~j = z.  \label{low_oldroydB3b}
\end{gather}
\end{subequations}

Thus, the dimensionless parameters that affect the dynamics of the interface can be identified as $\theta,~De_\theta,$ and $~Ec_{\theta}$.  In these shear dominated thin-film flows, the largest velocity gradients and therefore the largest stresses are in the $xz$ and $yz$ directions, corresponding to Eq. (\ref{low_oldroydB3a}). From the definition of the dimensionless parameters and Eq. (\ref{low_oldroydB3a}), we can observe for 
$\theta \ll 1$ that 
\begin{equation}
  \label{lowtheta_limit_result}
  \tau^p_{zj} \sim \theta~\frac{\partial u_j}{\partial z} +{\cal O}(\theta^2), ~~~j=x,y ,
\end{equation}
which suggests that polymer stress will not affect coalescence dynamics at leading order. The experiments  below agree well with this prediction. The result can be attributed to the significant dependence of the Deborah number on the contact angle, i.e. $De_\theta \sim \theta^3$. This relationship results from the characteristic velocity scaling similarly, $u_c \sim \theta^3$, which is a feature of capillarity-driven thin-film flows. The same result can be immediately seen by taking the $\theta \rightarrow 0$ limit of Eq. (\ref{eq_smalltheta_de_ec}), where $Ec$ becomes infinite and $De$ goes to zero. This is the Newtonian limit where there is no memory and the stress response is instantaneous.


\subsection{Experimental setup}
\subsubsection{Materials and Methods}
The experiments were performed using polyethylene oxide (PEO; Sigma-Aldrich; molecular weight $M_w=4 \times 10^6$ g/mol) dissolved in deionized water at three different concentrations: $c= 0.1$ wt\%, $0.5$ wt\%, and $1.0$ wt\%. In terms of the critical coil overlap concentration $c^*=1/(0.072 ~M_w^{0.65})$  above which the polymer coils may overlap one another \citep{tirtaatmadja2006drop}, these solutions correspond to $c/c^*=1.4,~7.0,$ and $14.1$, respectively. Note that the units of $c^*$ are g/mL and therefore the prefactor of the power law has units (mL mol$^{0.65}$)/(g$^{1.65}$). After the solutions were made, they were mixed with a magnetic stirrer for at least 24 hours. The surface tensions of the three solutions were measured using the pendant drop method and were $\gamma = [62.7 \pm 2.5,~62.8 \pm 1.9,~63.1 \pm 2.7]$ mN/m, respectively. The shear viscosities $\mu$ of the three solutions, at low shear rates, were measured using a rheometer (Anton Parr MCR 302e) and were $\mu = [2.37 \pm 0.24,~27.5 \pm 0.7, 263.4 \pm 8.8]$ mPa s. The relaxation times for the solutions were obtained from the literature \citep{dekker2022elasticity}, where they were measured from the thinning dynamics of the neck during pinch-off of a PEO drop.

We used glass microscope slides as the substrate for the experiments. They were cleaned by sequentially immersing and sonicating the slides for 15 minutes each in a surfactant solution, deionized water, ethanol, and acetone bath. A clean substrate is very important to prevent pinning of the triple line and to make sure that the two drops have the same contact angle upon contact and coalescence. The contact angle $\theta$ of the drops with the substrate at the moment of coalescence was within $7.5^{\circ}$ to $12^{\circ}$ ($0.13$ to $0.21~\textrm{radians}$). The drops were dispensed successively through a needle and were placed far enough apart to spread and reach small angles before contact. The angles of the drops were measured a posteriori from the experimental images to confirm symmetric coalescence. 

\subsubsection{Imaging and processing}
\begin{figure}
\begin{center}
\includegraphics[width=0.9\textwidth]{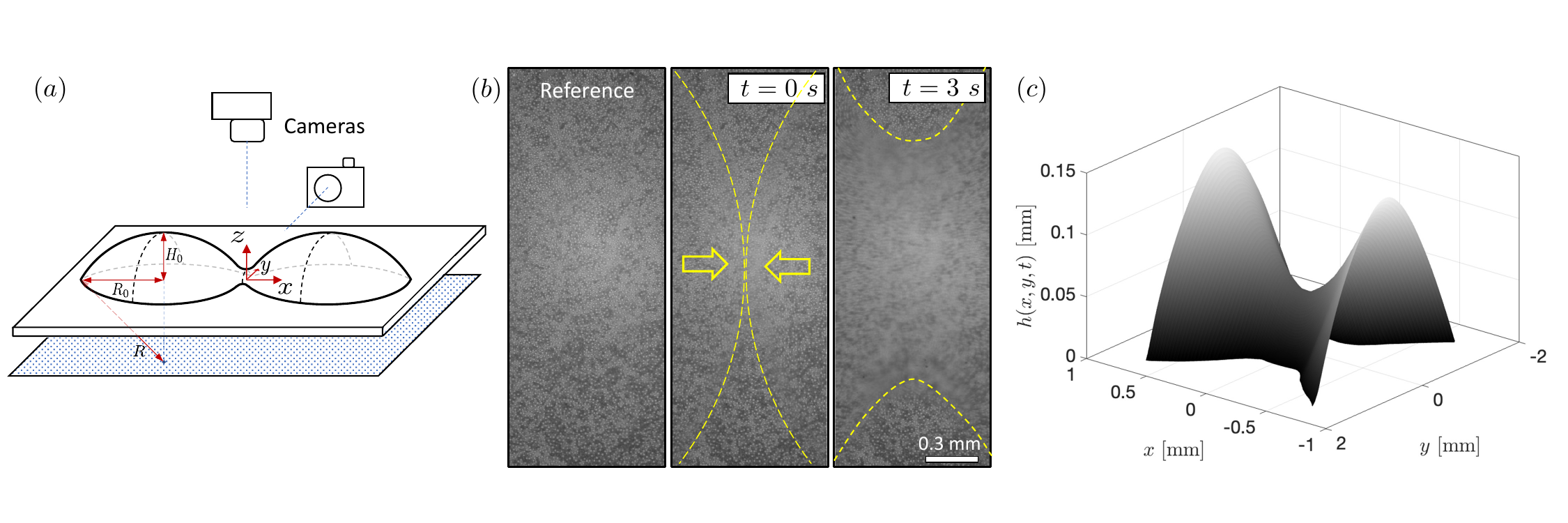}
\caption[Three-dimensional reconstruction of the shape of the interface using Free-Surface Synthetic Schlieren imaging.]{\label{ch6fig2} Three-dimensional reconstruction of the shape of the interface using Free-Surface Synthetic Schlieren imaging. (a) Schematic showing the experimental setup and the drop geometry. (b) Sequence of experimental images showing a reference frame taken before the drop appeared, and two time steps during the spreading and coalescing of 1 wt\% PEO drops. (c) The 3D reconstruction of the interface shape corresponding to the $t=1$ s after coalescence.  }
\end{center}
\end{figure}
In order to image the three-dimensional shape of the interface during coalescence, we used the Free-Surface Synthetic Schlieren (FS-SS) imaging \citep{moisy2009synthetic}. In this setup, a dot pattern is placed underneath the substrate that holds the drops and the imaging is done with a camera from the top (Phantom V7.3; 400 fps). Figure \ref{ch6fig2}(a) shows a schematic of the FS-SS setup. This technique utilizes the fact that the curved shape of the drop acts as a lens to distort the image of the dot pattern that is under the drop. By measuring the distortion, or the displacement field of the dot pattern compared to a reference image, we can calculate the shape of the interface that caused the distortion. Figure \ref{ch6fig2}(b) shows a reference image and two images at $t=0$ s and $t=3$ s during the coalescence of 1\% PEO drops. Notice that the latter two images show a distorted dot pattern relative to the reference image. Figure \ref{ch6fig2}(c) shows the reconstructed 3D shape of the interface from the experimental images. More information about how we calculate the interface profile from the images can be found in Appendix \ref{imageprocess-appD}.

\subsection{Results and discussion}
\subsubsection{Height of the bridge}
\begin{figure}
\begin{center}
\includegraphics[width=0.9\textwidth]{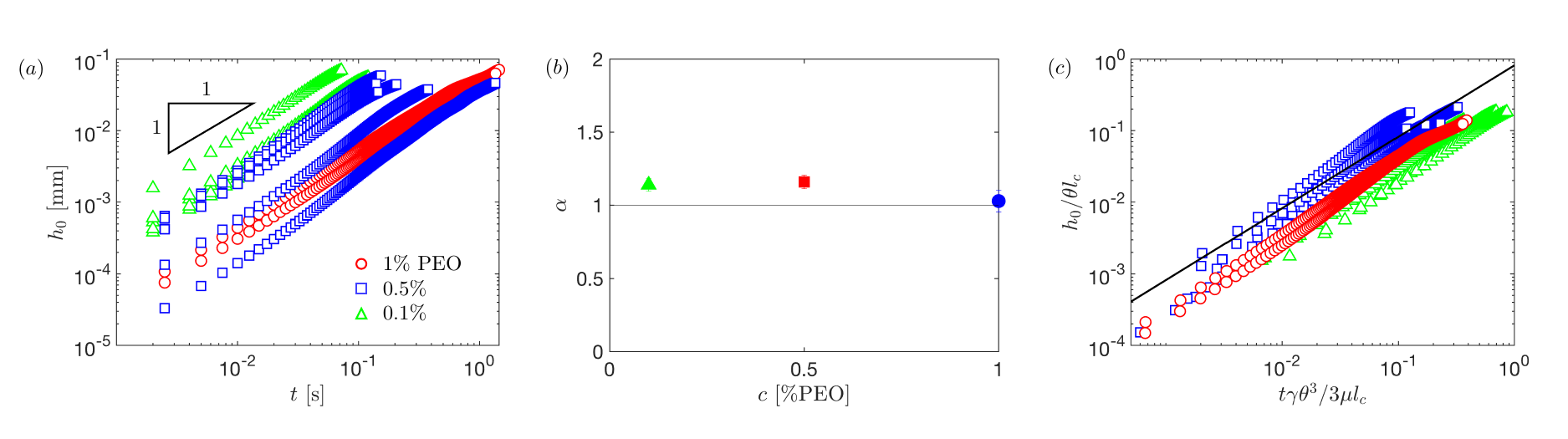}
\caption[The time evolution of the height of the interface $h_0(t)$ at the initial coalescence point.]{\label{ch6fig3} The time evolution of the height of the interface $h_0(t)$ at the initial coalescence point. (a) Raw data showing $h_0(t)$ from experiments using three different polymer concentrations, spanning the ranges $De_\theta = [0.002,~0.06]$ and $Ec_\theta = [0.07,~1.7]$. (b) Average power-law exponent $\alpha$ from fitting the data for the different polymer concentrations. (c) The $h_0$ versus $t$ data rescaled according to the Newtonian viscous scaling. Rescaling reasonably collapses the data, and the black line has a power-law exponent $\alpha=1$ and a prefactor $A=0.818$, predicted by the viscous theory. }
\end{center}
\end{figure}

In coalescence problems, analyzing the height of the bridge is a starting point for understanding the dynamics of the system \citep{hernandez2012symmetric, kaneelil2022three}. Note that the height of the bridge, $h_0(t)$, refers to the height of the interface at the point where the two drops initially made contact. Figure \ref{ch6fig3}(a) summarizes the time evolution of the height of the bridge for different polymer concentrations. Data from multiple experiments are shown for each polymer concentrations and span the ranges $De_\theta = [0.002,~0.06]$ and $Ec_\theta = [0.07,~1.7]$. Theory for the evolution of the height of the bridge for viscous Newtonian drops predicts a linear scaling with time,  $h_0(t) =vt= (A\theta u_c) t=A \frac{\gamma \theta^4}{3 \mu}t$, where $v$ is the velocity in the $z$ direction and $A \approx 0.818$ is a prefactor that can be determined uniquely \citep{hernandez2012symmetric}. Note that the variability in the $h_0(t)$ data for a given polymer concentration in Fig. \ref{ch6fig3}(a) arises from the difference in $\theta$ between experiments, which leads to varying vertical velocities of the interface. We fit the data using a power-law, $h_0(t) \propto t^\alpha$, and the results are shown in Fig. \ref{ch6fig3}(b). The average power-law exponent $\alpha$ for the 0.1\%, 0.5\%, and 1\% PEO cases were $1.14\pm0.04,~ 1.16\pm0.05$, and $1.03\pm0.08$, respectively. The coefficient of determination in all cases were above $R^2=0.99$.

We rescaled the height evolution data according to the Newtonian scaling, as shown in Fig. \ref{ch6fig3}(c), and observed a reasonable collapse of the data. The rescaled data is expected to have a slope of $A$, which is the slope of the black line plotted in Fig. \ref{ch6fig3}(c). Although the data shows reasonable collapse compared to raw data, there exists some variability. We believe that this variability originates from the error in experimental measurement of the contact angle of the drops, which gets amplified due to the $\theta^4$ dependence. We now reveal the shape of the interface and show that it is also in good  agreement with the Newtonian counterpart of this experiment.

\subsubsection{Shape of the interface}

\begin{figure}
\begin{center}
\includegraphics[width=0.9\textwidth]{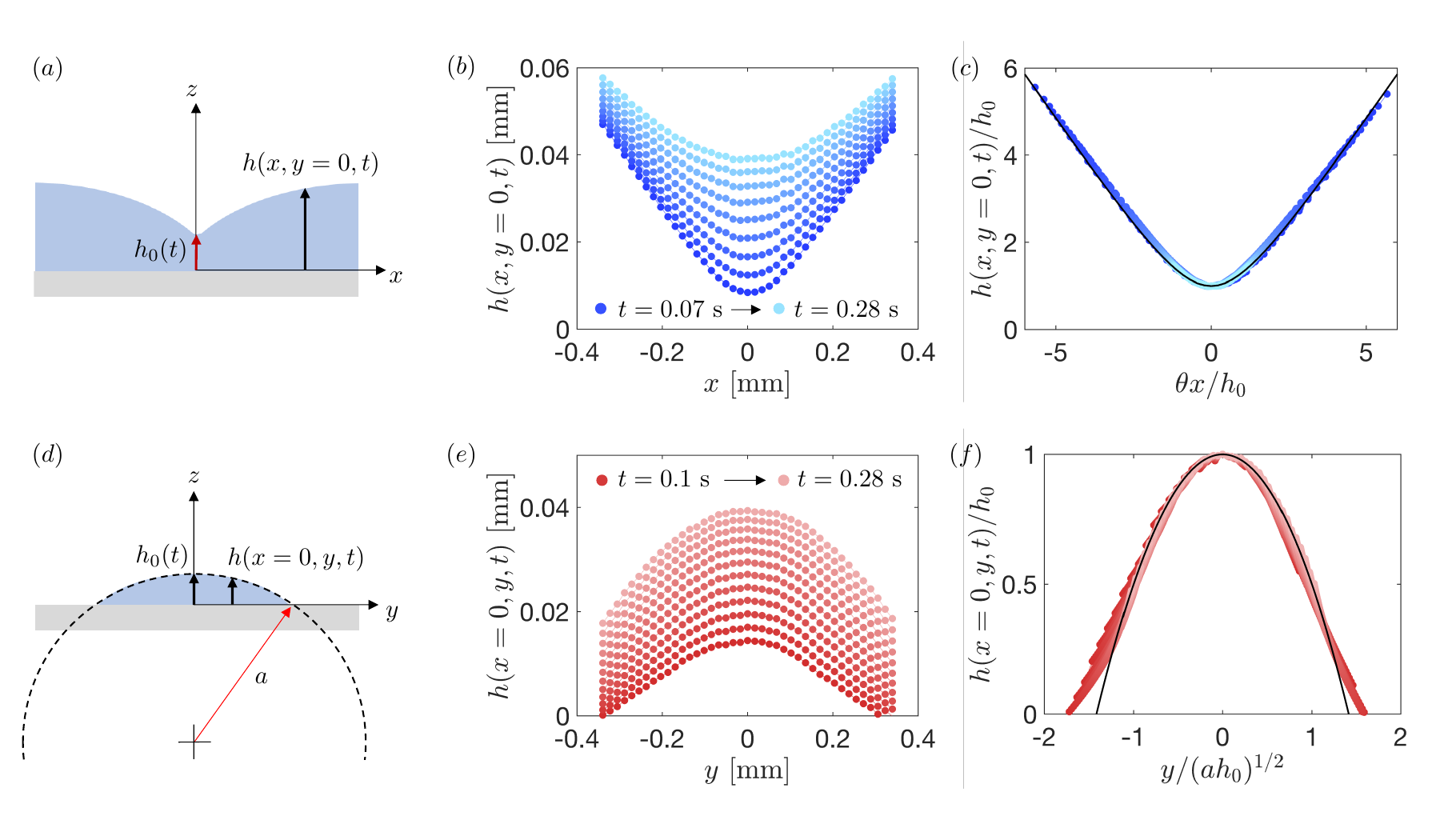}
\caption[The interface profiles along the $x$ and $y$ axes from the coalescence of 0.5 wt\% PEO drops with $\theta \approx 8.1^{\circ}$.]{\label{ch6fig4} The interface profiles along the $x$ and $y$ axes from the coalescence of 0.5\% PEO drops with $\theta \approx 8.1^{\circ}$, corresponding to $De_\theta = 0.009$ and $Ec_\theta = 0.74$. (a) Schematic of the interface in the $xz$-plane where the height $h_0(t)$ at the coalescence point is labelled. (b) Experimental data showing the dynamic shape of the interface in this plane. Notice that the darker colored markers correspond to earlier times and the lighter colored ones to the later times. Markers are connected by a faint line that is intended to only serve as a guide for the eyes. (c) The interface profiles rescaled with $h_0(t)$. The black line is the self-similar profile in the $xz$-plane. (d) Schematic of the interface in the $yz$-plane, where $a$ is the radius of a spherical cap. (e) Experimental data showing the dynamic shape of the interface in the $yz$-plane. (f) The rescaled interface profiles with $a=2.7$ mm. }
\end{center}
\end{figure}

During coalescence of sessile drops, the shape of the interface that forms near the coalescence point is three-dimensional and resembles a saddle \citep{kaneelil2022three}. Unlike the case of  coalescence of spherical drops, the three dimensionality arises from the presence of the contact line. In the $xz$ plane, the height at the coalescence point $h_0(t)$ will be the lowest point on the interface and the interface will slope upwards to form the edge of the drop. In the $yz$ plane, $h_0(t)$ will be the highest point on the interface and the interface will slope downwards on either side toward the contact points on the substrate (see Fig. \ref{ch6fig2}(a) and Fig. \ref{ch6fig4}(a, d)). Figure \ref{ch6fig4}(b) shows the dynamic shape of the interface in the $xz$ plane, $h(x,y=0,t)$, as two drops of 0.5 wt\% PEO coalesce. In the Newtonian case, the height profile in the $xz$ plane has a self-similar profile where both $h(x,y=0,t)$ and $x$ are rescaled by $h_0(t)$ \citep{hernandez2012symmetric}. The results in Fig. \ref{ch6fig4}(c) show that this self-similar scaling collapses the data well. 


Next, we present the interface profile at different times along the $yz$ plane (Figure \ref{ch6fig4}(e)). If the coalescence dynamics were similar to that of the Newtonian case in the thin-film regime, we would expect the shape of the interface along the $yz$ plane to be parabolic. The parabolic profile is a limit of a circular segment shape when $\theta \ll 1$ \citep{kaneelil2022three}. Figure \ref{ch6fig4}(f) shows the rescaled profiles in the $yz$ plane and the black line is $h(x=0,y,t)/h_0(t)=1- 1/2 \big(y/\sqrt{a h_0(t)}\big)^2$. Note that the parameter $a$ that appears in the rescaling is a geometric parameter that captures the outer length scale of the system \citep{kaneelil2022three}. The data collapses well and shows good agreement with the parabolic profile. 

\begin{figure}
\begin{center}
\includegraphics[width=0.7\textwidth]{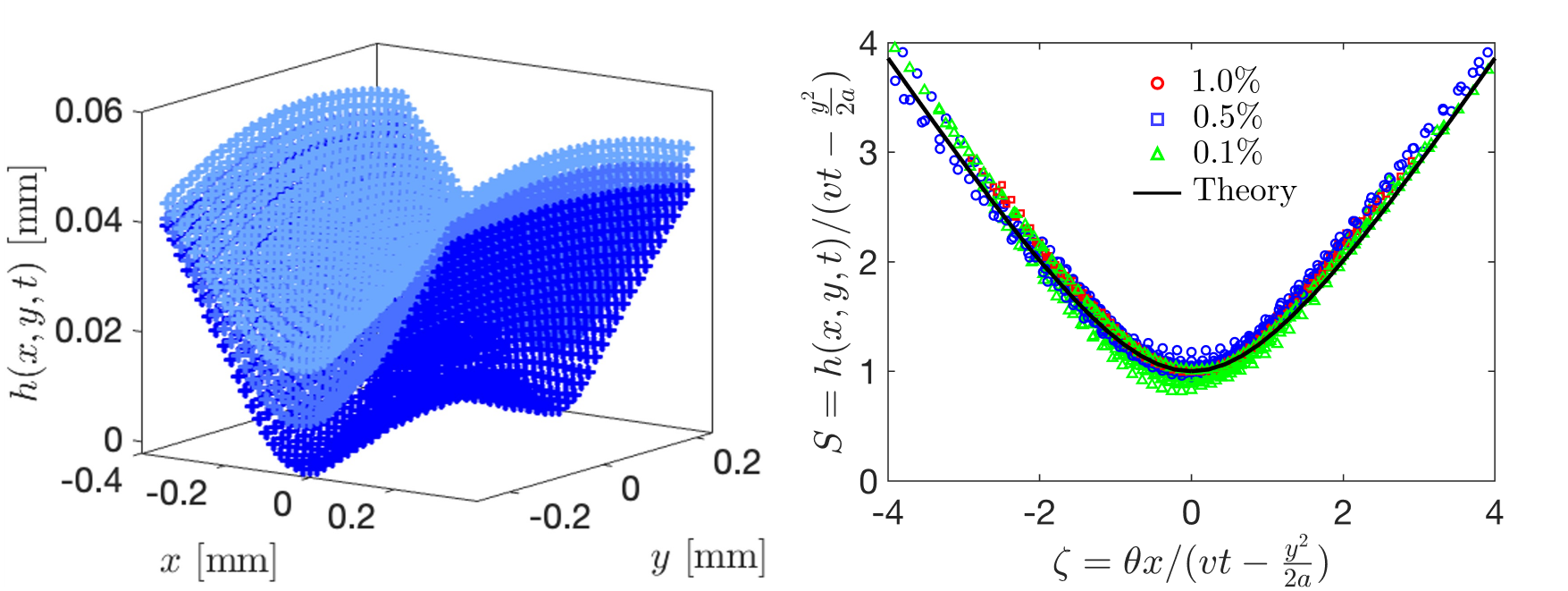}
\caption{\label{ch6fig5} The Newtonian three-dimensional self-similarity also describes the coalescence of semi-dilute polymeric drops at small $\theta$. (a) Experimental data from the coalescence of 0.5\% PEO drops with $\theta \approx 8.1^{\circ}$, corresponding to $De_\theta = 0.009$ and $Ec_\theta = 0.74$, showing the three-dimensional shape of the interface near the coalescence point at early times ($t= 0.05,~0.15,~0.22$ s). The darker colored markers correspond to earlier times and the lighter colored ones to the later times. (b) Experimental data from the coalescence of 0.1\%, 0.5\%, and 1.0\% PEO drops at 4 different times and 3 different $yz$-planes (total of 36 curves) rescaled according to the similarity solution. The rescaled data collapses onto the universal self-similar curve (black line). }
\end{center}
\end{figure}

Thus, we have shown that the spatiotemporal dynamics of the interface in the $xz$ and $yz$ planes agree with that expected for Newtonian coalescence, which suggests that the polymer has negligible effect on coalescence. We now show that the dynamic shape of the three-dimensional interface near the coalescence point can therefore be mapped to a 2D self-similar curve \citep{kaneelil2022three}. Figure \ref{ch6fig5}(a) shows the time evolution of the 3D interface for an experiment with 0.5 wt\% PEO drops. Rescaling the height as $S(\zeta) = h(x,y,t)/[vt-y^2/(2a)]$ where the similarity variable is $\zeta=\theta x/[vt-y^2/(2a)]$ collapses the data onto the self-similar shape, as shown in Fig. \ref{ch6fig5}(b).

In the small contact angle  limit, $\theta \ll 1$, we showed experimentally that the coalescence dynamics of semi-dilute polymer solutions is unaffected by polymers, as predicted by our scaling analysis. However, we note that this is not a generic result for all thin-film problems. In the classical Landau-Levich-Derjaguin problem of drawing an object out of a bath of liquid and analyzing the thickness of the liquid film left behind on the object, studies have shown that the thickness of the film is smaller for a weakly viscoelastic liquid relative to the Newtonian result \citep{datt2022thin}, and larger when the viscoelastic effects are increased \citep{lee2002study}. The lack of an effect of the polymer  that we observe can be attributed to the strong and unique dependence of the Deborah number on the contact angle, $De_{\theta} \sim \theta^3$, as discussed in section \ref{seclowangletheory}. Notice that the scaling arises from the definition of the characteristic velocity in this problem $u_c = \gamma \theta^3/(3\mu)$, which is a natural scale for capillarity-driven thin-film flows. Thus, we would expect our result that thin-film flow dynamics is unaffected by polymers to hold for problems that are internally driven by capillarity and not externally driven, e.g., boundary driven, as in the case of the Landau-Levich-Derjaguin problem. This feature can be leveraged in applications involving thin-film flows of polymeric liquids, where it is advantageous to minimize elastic effects and attain outcomes similar to those achieved with Newtonian fluids.

\section{The large contact angle limit}
\label{sechighangle}
We perform 2D numerical simulations of sessile drop coalescence using the open source partial differential equation solver Basilisk C \citep{popinet2009surftension,popinet2015basiliskquadtree} using viscoelastic constitutive equations \citep{turkoz2018axisymmetric,lopez2019adaptive}, along with some experiments to validate the numerical results in the large contact angle limit. We use the Oldroyd-B model to study the viscoelastic effects, since it is one the simplest models, and will show that it is sufficient to explain the regime of interest in this work. 

While our experimental results show that coalescence is unaffected by polymer at small $\theta$, it has been shown experimentally to affect coalescence dynamics at larger $\theta$ and when inertia is important \citep{dekker2022elasticity,varma2021coalescence}. Note that the characteristic scales that describes the flow dynamics are expected to be different here since inertia will not be negligible, whereas the $\theta \ll 1$ approximation in the previous section naturally led to viscously dominated flows. The goal with the simulations is therefore to further probe the inertially dominated large $\theta$ limit to reveal the underlying mechanism that is leading to a departure from the Newtonian coalescence dynamics.

\subsection{Dimensionless equations and numerical simulations}
We rescale the governing equations from section (\ref{goveq}) to make them dimensionless using the following characteristic scales that are relevant for inertial capillary-driven flows: $u=\Tilde{u}/U,~t=\Tilde{t}/\sqrt{\rho H^{3}/\gamma},~y=\Tilde{y}/H,$ and $x=\Tilde{x}/H$, where the typical velocity $U$ is taken to be the capillary velocity $U=H/t_c = (\gamma/\rho H)^{1/2}$ \citep{eggers1997ReviewModPhys}, and $H$ is the height at the center of the spherical cap shaped drop. Notice that the characteristic scales are different from those in section (\ref{seclowangletheory}), which was the viscously dominated thin-film regime. Non-dimensionalizing Eq. (\ref{eqch6momentum}) gives
\begin{equation}
    \frac{\textrm{D} \pmb{u}}{\textrm{D} t} = -\nabla p + Oh \nabla^2 \pmb{u} + \nabla \cdot \pmb{\tau}^p, 
    \label{eqch6_dimlessMomen}
\end{equation}
where the Ohnesorge number is $Oh = \mu_s/\sqrt{\rho \gamma H}$, the pressure scale is chosen as $p_c = \gamma/H$, and the scale for the polymer stress was set to be the same as the pressure scale in order for polymer effects to appear in the first order. Non-dimensionalizing Eq. (\ref{oldroydBgeneral}) gives 
\begin{subequations}
\label{high_oldroydB}
\begin{gather}
\pmb{\tau}^p=Ec(\pmb{A}-\pmb{I}), \label{eqch6oldroydDimless2}
\\
\overset{\mathtt{\nabla}}{\pmb{A}}=-\frac{1}{De}(\pmb{A}-\pmb{I}), \label{eqch6oldroydDimless3}
\\
De~\overset{\mathtt{\nabla}}{\pmb{\tau}^p} + \pmb{\tau}^p = 2~ Ec De~ \pmb{E}, \label{eqch6oldroydDimless1}
\end{gather}
\end{subequations}
where the Deborah number and the elastocapillary number are defined as 
\begin{equation}
    De = \frac{\lambda}{\sqrt{\rho H^{3}/\gamma}}, ~~~\textrm{and}~~~ Ec=\frac{GH}{\gamma}.
    \label{eq_de_ec}
\end{equation}

Thus, the dimensionless parameters that affect the dynamics of the interface can be identified as $Oh,~De,$ and $~Ec$. In Basilisk C, the implementation of the Oldroyd-B model uses the so-called log conformation technique, where the logarithm of the conformation tensor $\pmb{A}$ is calculated \citep{fattal2004constitutive,hao2007simulation,turkoz2018axisymmetric,lopez2019adaptive}. We used a modified version of the Oldroyd-B implementation that specifically identifies the modulus of the polymer instead of the solvent viscosity as one of the parameters, in addition to the relaxation time \citep{dixit2024viscoelastic}.

\begin{figure}
\begin{center}
\includegraphics[width=0.6\textwidth]{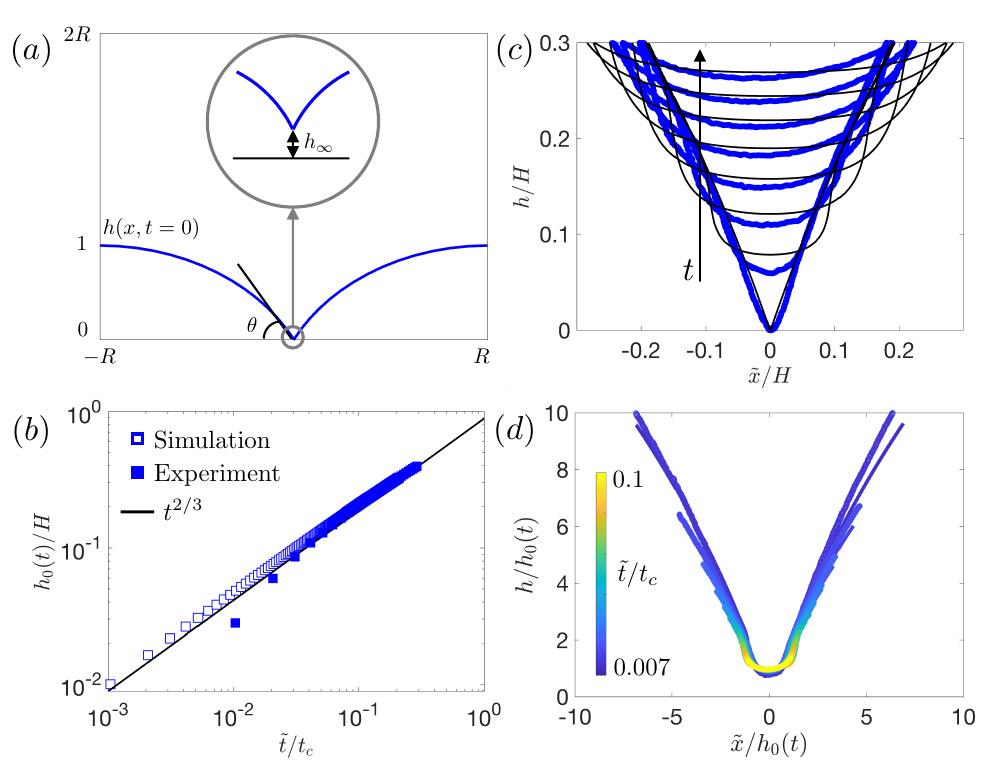}
\caption{\label{expsimcomparison} Simulation setup and comparison between simulation and experiments. 
(a) The initial shape of the interface $h(x,t=0)$, which follows Eq. (\ref{eqch6geometry}), for a case with $\theta=1.1$ and $h_{\infty}=0.008$ is shown in the numerical domain of length $L_0=2R$. The inset shows the initial film thickness $h_{\infty}$ at the coalescence point more clearly. 
(b) The temporal evolution of the height $h_0(t)$ at the coalescence point. The filled in markers are data from experiments and the hollow markers are data from simulations for $Oh = 0.004,~\theta=1.1,~De=0$, $Ec=0$.
(c) The experimental results (blue markers) overlaid on top of the simulation results (black lines) for $t \approx [0,~0.15]$.
(d) The shapes of the interface from experiment (dots) and simulation (lines) collapse on to a self-similar profile when rescaled with $h_0(t)$.
}
\end{center}
\end{figure}

The 2D simulation is initialized in a square box of length $L_0=2R$ with the shape of the interface defined as two symmetric circular segments, with all length rescaled by $H$, as follows:
\begin{equation}
\label{eqch6geometry}
\begin{split}
    h(x,t=0) = {\cal H} (-x) \bigg[ \bigg( \frac{1}{(1-\textrm{cos}(\theta))^2} - (x+R)^2 \bigg)^{1/2} - \frac{1}{1-\textrm{cos}(\theta)} + 1\bigg] \\ + {\cal H} (x) \bigg[ \bigg( \frac{1}{(1-\textrm{cos}(\theta))^2} - (x-R)^2 \bigg)^{1/2} - \frac{1}{1-\textrm{cos}(\theta)} + 1\bigg] + h_{\infty}.
\end{split}
\end{equation}
Here, ${\cal H}$ is the Heaviside step function, $\theta$ is the angle that the circular segments make with the horizontal, $R=\big[ \frac{1}{(1-\textrm{cos}(\theta))^2} - \big(\frac{\textrm{cos}(\theta)}{1-\textrm{cos}(\theta)} \big)^2\big]^{1/2}$ is the rescaled radius of the base of the spherical cap, and $h_\infty$ is the arbitrarily small film thickness at the point where the two circular segments meet. Notice that $h(x,t=0)$ is fully specified by a single parameter $\theta$. This definition of the interface shape eliminates the presence of a contact line. Figure \ref{expsimcomparison}(a) shows the simulation domain and the initial shape of the interface for a case with $\theta=1.1~\textrm{radians}$ and $h_{\infty}=0.008$ (this value of $h_{\infty}$ is kept constant for all simulations shown here and was verified to not affect the results). We impose a no-slip boundary condition on the bottom of the domain, symmetry conditions on the left and right, and outflow on the top. The self-similar dynamics of the interface that we are interested in take place at very early times of coalescence, when the height of the bridge is smaller than the macroscopic length scale of the drop $h_0 \ll H$, and should be unaffected by the symmetry conditions along the sides of the domain and other boundary effects. 

We use an adaptive quadtree grid where the maximum level of refinement $L$ is specified, and the total number of grid points in the vertical or horizontal direction is $2^L$ \citep{popinet2015basiliskquadtree,van2022fourth,mostert2022high}. We varied the grid refinement, as shown in the Appendix \ref{appAGrid}, to ensure that the results are grid independent. We use $L=11$ for all the simulations shown here, which corresponds to five grid points within a height of $h_{\infty}$.

\subsection{Comparing experiments and simulations}
We start by comparing the results between experiments and simulations for a Newtonian case with $Oh = 0.004,~\theta=1.1,~De=0$, and $Ec=0$, which corresponds to an experiment with water drops. Note that these large $\theta$ experiments are imaged from the side (see Fig. \ref{ch6fig2}(a)) using a high speed camera (Phantom V2012; 29,000 fps). The large contact angles were achieved in experiments with glass microscope slides that were simply rinsed with water and dried with compressed air.

The results in Fig. \ref{expsimcomparison}(b) show the comparison of $h_0/H$ as a function of time $\tilde{t}/t_c$ between the simulations and experiments. There is excellent agreement between the rescaled simulated and experimental data. The slight deviation from the simulation results seen in the experimental data at very early times may be due to the error in measurement of the small $h_0(t)$ values when it is close to the resolution of the camera. 

In the Newtonian coalescence of sessile drops at large $\theta$ and low $Oh$, the dynamics of the height at the coalescence point follows $h_0(t)/H = c_0 (\tilde{t}/t_c)^{2/3}$. A value of $c_0=0.89$ is used for the solid black line in Fig. \ref{expsimcomparison}(b), similar to previously reported results \citep{eddi2013influence}. Both the simulated and the experimental data in Fig. \ref{expsimcomparison}(b) are in agreement with this scaling.

Figure \ref{expsimcomparison}(c) shows the dynamic shape of the interface from the experiments with water (blue markers) overlaid on top of the results from the simulation (black lines) in rescaled units $\tilde{x}/H$ and $h/H$. One artifact of the 2D simulations is the more pronounced capillary waves seen at the interface leading to a wider interface shape than those observed in the inherently 3D experiments \citep{keller2000merging,eddi2008wave,eddi2013influence}. Despite this slight discrepancy, the simulations and the experiments are in good agreement at early times of coalescence when there is a clear separation of scales. Figure \ref{expsimcomparison}(d) shows the shape of the interface obtained from both simulations (lines) and experiments (markers) rescaled according to the Newtonian self-similar scaling, $h/h_0(t)$, and $x/h_0(t)$, which shows reasonable collapse.


We now use the numerical simulations to further study the dynamics of viscoelastic coalescence over a wide range of $De$ and $Ec$ in order to see how the coalescence dynamics deviate from the Newtonian behavior as a function of these dimensionless parameters.

\begin{figure}
\begin{center}
\includegraphics[width=1\textwidth]{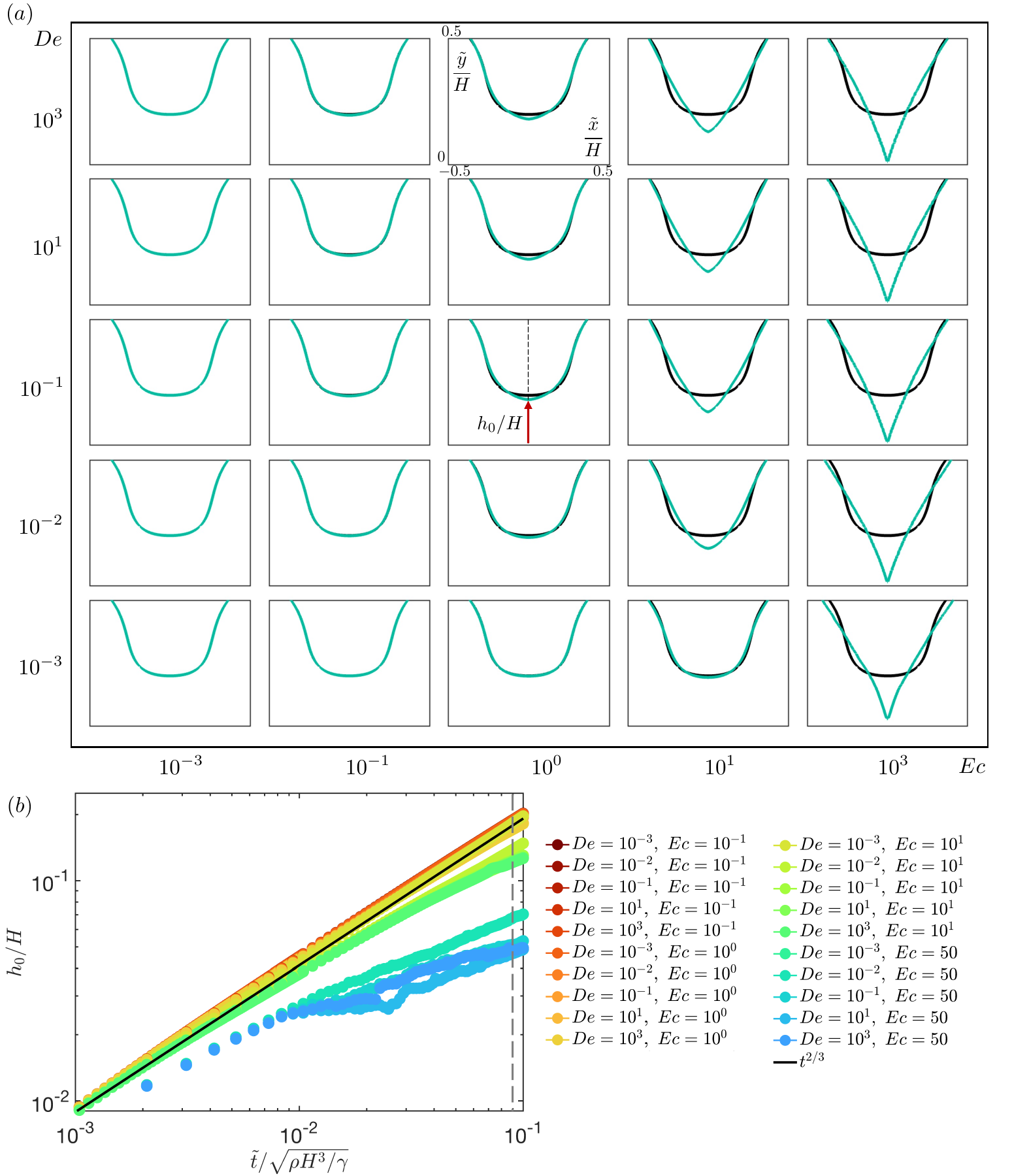}
\caption{\label{ch6_De-B_sweep} Effect of $De$ and $Ec$ on the shape of the interface. (a) Snapshots of the interface at $t = 0.09$ for $Oh=0.01$ and $\theta=1.1$ for various $De$ and $Ec$ combinations. Note that the solid black line correspond to the Newtonian limit with $De=Ec=0$, and the colored lines correspond to the $De$ and $Ec$ shown on the axes. (b) The height at the coalescence point, $h_0$, is plotted as a function of time for various $De$ and $Ec$. The solid black line corresponds to $t^{2/3}$ while the dashed gray line marks $t=0.09$, the time at which the shapes in panel (a) are shown.}
\end{center}
\end{figure}

\subsection{The effect of $De$ and $Ec$ in the coalescence dynamics}

The conspicuous effect of the polymers on the coalescence dynamics of large $\theta$ drops is seen in the shape of the interface, as shown in Fig. \ref{ch6fig1}(b)-(c), where the curvature of the interface seems to be high in the presence of polymers. We now aim to understand the extent of this effect as a function of $De$ and $Ec$. In all the following results, we keep the values of Ohnesorge number fixed at $Oh=0.01$, ensuring that inertial-capillary effects dominate over viscous effects, the contact angle is $\theta=1.1$, and focus solely on the physics as a function of $De$ and $Ec$. 

Figure \ref{ch6_De-B_sweep}(a) shows snapshots of the shape of the interface during coalescence at $\tilde{t}/t_c=0.09$ for $Oh=0.01$ and $\theta=1.1$ for various values of $De$ and $Ec$. The colored curve represents the shape of the interface at the given $De$ and $Ec$, and the black curve represents the Newtonian case with $De=Ec=0$. When $De$ and $Ec$ are small, the shape of the interface overlaps with that of the Newtonian case. This limit corresponds to the case of dilute polymer solutions where the coalescence seems to be unaffected by polymers. As $De$ and $Ec$ increases, we see that the shape of the interface starts to change from the Newtonian response. The shape of the interface and therefore the viscoelastic effects seems to be more sensitive to $Ec$ and shows almost no change from the Newtonian results when $Ec$ is sufficiently small.

The results we see in the phase plane in Fig. \ref{ch6_De-B_sweep}(a) can be understood from the various limits of Eq. (\ref{high_oldroydB}). First, for small $De$, i.e. $De\rightarrow 0$, we see from Eq. (\ref{eqch6oldroydDimless1}) that $\pmb{\tau}^p \approx 2 Ec~De~\pmb{E}$. If $Ec~De$ is also small, i.e. $Ec \rightarrow 0$, then we have $\pmb{\tau}^p \rightarrow 0$, which is the Newtonian limit. If $Ec~De$ is finite or large, this scenario corresponds to the case where the total viscous response of the fluid is altered by the polymeric viscosity such that the deviatoric stress becomes $(Oh+Ec~De)\nabla^2 \pmb{u}$. On the other hand when $Ec \rightarrow 0$, we have $\pmb{\tau}^p \rightarrow \pmb{0}$ independent of $De$, and the dynamics is purely Newtonian as seen in Fig. \ref{ch6_De-B_sweep}(a).

As $De \rightarrow \infty$, we have $\overset{\mathtt{\nabla}}{\pmb{\tau}^p}=2Ec\pmb{E}$ where the magnitude of $Ec$ dictates whether the response is Newtonian or elastic. And finally as $Ec \rightarrow \infty$, we have $\pmb{\tau}^p \rightarrow \infty$ and $\pmb{E} \rightarrow \pmb{0}$ corresponding to minimal deformation, as clearly seen by the undeformed interfaces in Fig. \ref{ch6_De-B_sweep}(a). We note that the actual results seen in experiments for $De,~Ec \rightarrow \infty$ might be quantitatively different from those seen in the simulations here since the Oldroyd-B model does not capture the finite extensibility of polymer chains.

The dynamics of the height of the bridge, $h_0/H$, is shown in Fig. \ref{ch6_De-B_sweep}(b) for several $De-Ec$ combinations. Note that the data for $Ec=10^3$ isn't shown since the interface was effectively stationary. When $De$ and $Ec$ are sufficiently small, the temporal evolution follows the scaling $t^{2/3}$ and the data is collapsed quite well with the Newtonian and inertial rescaling factors with the same prefactor of $c_0=0.89$ (black line in Fig. \ref{ch6_De-B_sweep}(b)). But, as the dimensionless parameters are increased, with $De \approx 10^{-3}$ and $Ec \approx 10^1$ and larger, a decrease from the $t^{2/3}$ scaling is observed.  Such a decrease in the scaling exponent as a function of the polymer concentration has been previously reported \citep{varma2021coalescence,varma2022elasticity,rostami2025coalescence}. However, we note that the scaling exponent in our study and in the results from \cite{dekker2022elasticity} goes lower than 0.5, perhaps since the way the drops come into contact is different than the droplet spreading method where the drops impact the substrate, spread, and contact each other \citep{varma2022elasticity}. We also point out that the departure from the Newtonian scaling can be seen in this study simply from the Oldroyd-B model without any need to consider shear thinning and finite extensibility behavior. We now probe the evolution of the stress fields during coalescence to gain insight into this departure.

\begin{figure}
\begin{center}
\includegraphics[width=0.8\textwidth]{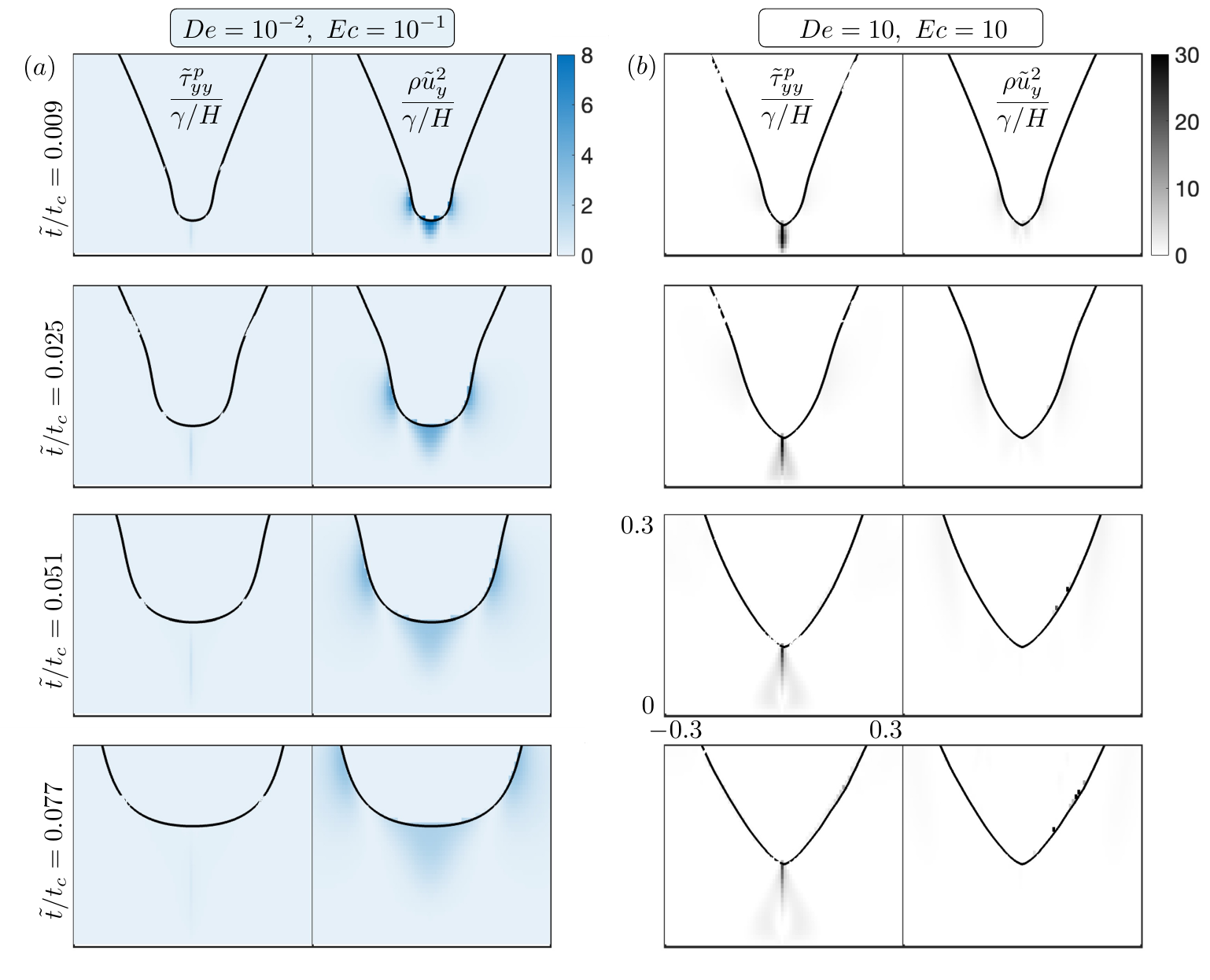}
\caption{\label{stress_evolv} The evolution of the polymeric and inertial stress fields for simulations with $Oh=0.01$ and $\theta=1.1$. The dimensionless polymeric stress field $\Tilde{\pmb{\tau}}^p_{yy}/(\gamma/H)$ is plotted on the left column and the dimensionless inertial stress $\rho \Tilde{u}_y^2/(\gamma/H)$ is plotted on the right, for four time steps. (a) A case with $De=10^{-2},~Ec=10^{-1}$ is observed to be dominated by inertia. (b) A case with $De=10,~Ec=10$ exhibits a much sharper interface shape and has a larger polymeric stress field. Note that the region of large polymeric stress is localized to the middle where the drops initially made contact. }
\end{center}
\end{figure}

The larger curvature seen in the shape of the interface during viscoelastic coalescence (see Appendix \ref{curvature-appc}) suggests an increase in stress, presumably due to the contribution from $\pmb{\tau}_p$, which we show is highly localized near the coalescence point. Figure \ref{stress_evolv}(a) shows the dimensionless polymeric stress field (left column) and the dimensionless inertial stress field (right column) for $De=10^{-2}$ and $Ec=10^{-1}$, while Fig. \ref{stress_evolv}(b) shows the same for $De=10$ and $Ec=10$. Note that we consider locally averaged stress fields, as explained in Appendix \ref{avestress-appB}, which is also independent of the grid size. In the former case, fluid inertia dominates and the effect of the polymeric stress is relatively minimal. In the latter case, however, polymeric stress dominates and is about an order of magnitude higher than the inertial stress. The region where this large polymeric stress occurs is also localized to near the middle because of the highly extensional flow created by the rising interface. Consequently, the larger stress must be balanced by a local increase in the pressure field, leading to a sharper interface profile. 

Such a sharp shape of the interface was also observed at the rear end of gas bubbles rising in viscoelastic liquids, where the shape change was also attributed to locally large extensional stresses \citep{astarita1965motion,pilz2007critical,fraggedakis2016velocity}. These local extensional stresses caused by extended polymer molecules are generally present downstream of stagnation points and can be visualized in experiments using birefringence \citep{farrell1978observation,cressely1979lines,harlen1990high}. Due to this local nature, large elastic stresses only occur close to these \textit{strands} and the resulting flow can be modelled as a line distribution of forces in a Newtonian background fluid \citep{harlen1990high}. In our study, the vertical rise of the interface during coalescence sets up a stagnation point at the substrate directly below the smallest height of the interface, namely $h_0$. At sufficiently high $De$ and $Ec$, we expect the extensional flow resulting from coalescence to generate these \textit{strands} of high elastic stresses, as seen in Fig. \ref{stress_evolv}(b), and therefore alter the dynamics of the interface. We can probe the temporal evolution of the maximum values of the various stress fields to understand how the elastic stress develops as a function of our dimensionless parameters.

\begin{figure}
\begin{center}
\includegraphics[width=0.85\textwidth]{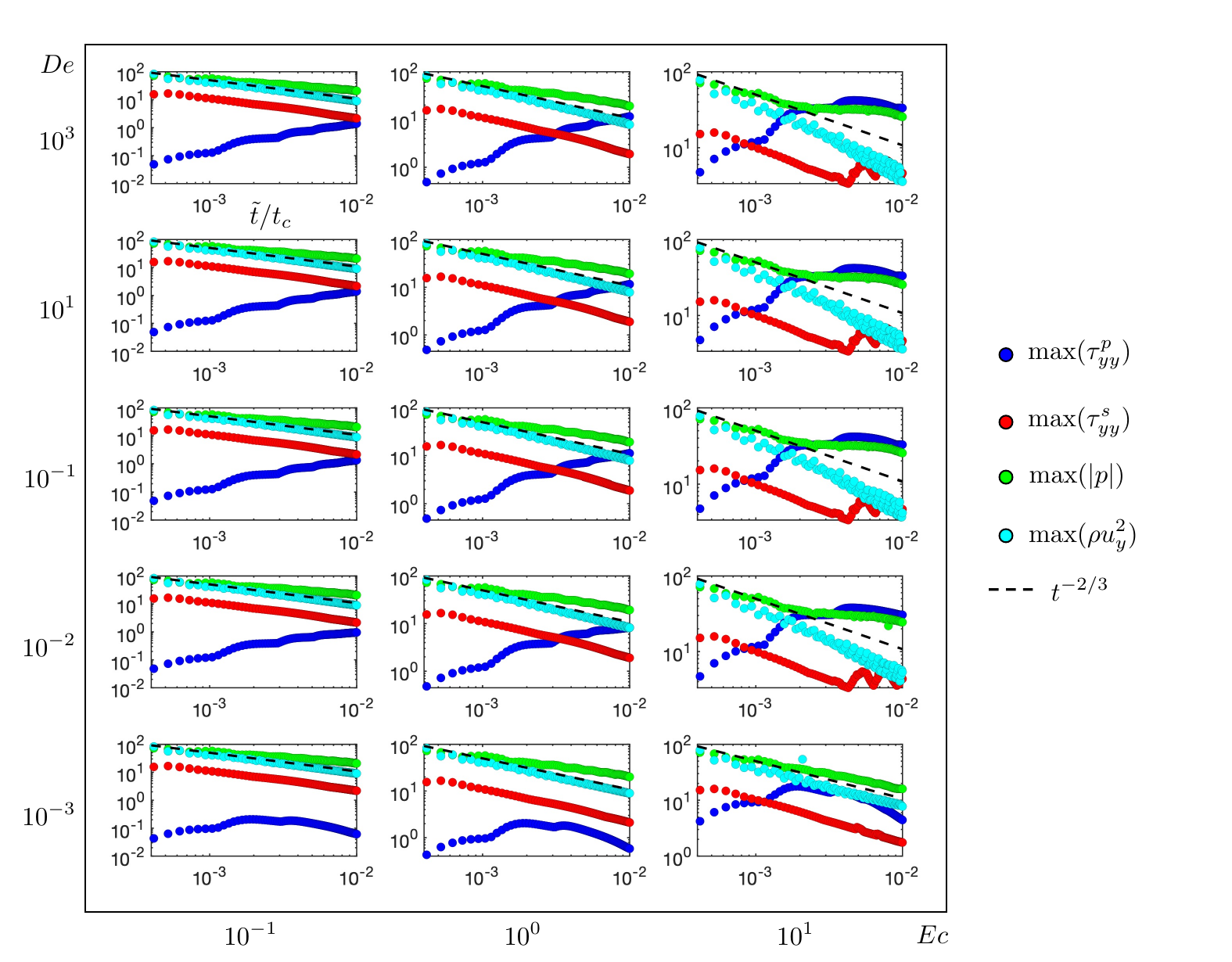}
\caption{\label{stress_phase_diag} The maximum values of various stress fields as a function of $\Tilde{t}/t_c$ during the early times of coalescence for a wide range of $De$ and $Ec$. The blue marker represents max($\pmb{\tau}^p_{yy}$), red is max($\pmb{\tau}^{s}_{yy}$), cyan is max($\rho u_y^2$), and green is max($|p|$). Note that all stresses are rescaled by $\gamma/H$. The dashed black line represent $h_0^{-1} \propto t^{-2/3}$ which is the rate at which capillary and inertial stress decays in the absence of polymers. }
\end{center}
\end{figure}

Figure \ref{stress_phase_diag} shows the evolution of the \textit{maximum} value of the stress fields at early times of coalescence for various $De$ and $Ec$. We keep $Oh=0.01$ and $\theta=1.1$ as before. The blue marker represents max($\pmb{\tau}^p_{yy}$), red is max($\pmb{\tau}^{s}_{yy}$), cyan is max($\rho u_y^2$), and green is max($|p|$). 
Note that the magnitude of pressure is taken since the pressure in the droplet phase will be negative during coalescence due to the reference pressure outside the drop being initially zero. Firstly, note that at very early times, the maximum value of pressure is independent of $De$ and $Ec$ since the curvature of the interface at the very initial moments of coalescence will likely only be a function of the contact angle $\theta$. Note also that the pressure here corresponds to the net stress in the system which is the same as the capillary stress since the interface shape reflects the state of stress. Once coalescence begins, the extent of how the other stresses balance pressure, or capillary stress, tells us about the dynamics of the interface. Quite remarkably, we see that at the very early moments of coalescence, the dominant balance is between capillary stress and inertia. Note that this will be true as long as $Oh$ is small and inertia dominates over viscous stress. When $De$ and $Ec$ are sufficiently small, inertial stresses primarily counterbalance the capillary stress throughout most of the coalescence process. However, as $De$ and $Ec$ increase, polymeric stress rapidly increases to become the predominant stress balancing capillary stress.

 In the initial stages of coalescence before the polymeric stress takes over, the capillary stress and the balancing inertial stress are reasonably expected to decay independent of $De$ and $Ec$ as $h_0^{-1} \propto t^{-2/3}$ (black dashed line in Fig. \ref{stress_phase_diag}). As polymeric stress increases to become the dominant contribution, the rate of decay of the capillary stress decreases, shown by the departure from the $t^{-2/3}$ behavior in Fig. \ref{stress_evolv} (see green data points). This decrease in the rate of decay of the capillary stress directly leads to a slower growth of $h_0$. Furthermore, since the curvature of the interface is tied to the stress field, a greater curvature is observed in the shape of the interface (see Figs. \ref{ch6_De-B_sweep} and \ref{stress_evolv}). This increase in the polymeric stress, which is spatially local in nature, and the subsequent slow down in the rate of decay of the capillary stress, is what causes the coalescence dynamics to deviate from the inertial and Newtonian behavior.

Interestingly, near $Ec \approx 10$ when the departure from the Newtonian dynamics is observed, the polymeric stress still requires a finite amount of time to grow and become the dominant contribution, allowing for the capillary and the inertial stresses to set the early time dynamical behavior. This finite timescale of the growth of the polymer stress is what allows the Newtonian scaling of the bridge height $h_0(t) \sim t^{2/3}$ to persist. However, we see that the dynamics slowly deviate from the $t^{2/3}$ scaling over time (see Fig. \ref{ch6_De-B_sweep}(b)) once the polymeric stress has become the dominant contribution.  While the departure from the Newtonian dynamics of the temporal evolution of the bridge is subtle and gradual, the effect on the shape of the interface is effectively instantaneous, since the shape of the interface must directly reflect the instantaneous state of stress. It is also more noticeable because the polymeric stress is highly localized (see Fig. \ref{stress_evolv}). This is clearly observed when analyzing the temporal evolution of the curvature of the interface (see Appendix \ref{curvature-appc}), which shows that the curvature for the viscoelastic case is much larger than its Newtonian counterpart whereas the temporal scaling remains the same.

Increasing $Ec$ seems to decrease the timescale of the growth of the polymer stress, leading to an earlier transition to polymeric stress dominance and allowing the polymeric stress to reach a greater magnitude. As a result, the interface shape is altered more significantly with larger $Ec$. Increasing $De$ increases the timescale over which the polymeric stress starts to decay and essentially dictates whether or not the transition to polymeric stress dominance occurs during the early stages of coalescence. Thus, we show that elasticity affects coalescence dynamics only at sufficiently large $De$ and $Ec$ due to the extensional flow giving rise to locally large elastic stress \textit{strands}. We believe that more experimental and theoretical studies are necessary to fully understand this complicated process of viscoelastic coalescence and reveal the self-similar dynamics which must account for the transition we see as a function of $De$ and $Ec$. Future work may also probe the effect of varying $Oh$ and $\theta$ and consider more sophisticated viscoelastic constitutive laws.

\section{Conclusions}
The coalescence of polymer drops was explored using experiments in the context of small $\theta$ and using experiments and direct numerical simulations in the context of large $\theta$. For the small $\theta$ case, we imaged the 3D shape of the interface during coalescence using Free-Surface Synthetic Schlieren and found that the coalescence dynamics is the same as that of viscous Newtonian drops. The negligible effect of the polymer on coalescence was attributed to the low Deborah number of the system, which scales with $\theta$ as $De \sim \theta^3$ in the small $\theta$ limit. When $\theta \ll 1$, we showed that the effect of the polymers only appear at ${\cal O}(\theta)$ and the coalescence dynamics simply follows the Newtonian self-similar dynamics for the thin-film regime. This result is important because viscoelasticity is known to affect the thin-film dynamics in other problems, such as the Landau-Levich-Derjaguin problem. Yet, our results show that coalescence is a scenario where the elastic effects can be effectively disregarded within the thin-film flow regime.

In the large $\theta$ case where inertia dominates the dynamics, coalescence is altered by the presence of polymers. We used direct numerical simulations with the Oldroyd-B model to study this limit. The two parameters that captures the effect of viscoelasticity were the Deborah number $De$ and the elastocapillary number $Ec$. As $De$ and $Ec$ are increased, we showed that the polymeric stress becomes the dominant stress contribution balancing capillary pressure. Revealing the stress fields inside the drop and the temporal evolution of the various stress fields, we provide more insight into how the coalescence dynamics starts to deviate from the classical Newtonian dynamics for viscoelastic drops.

\vspace{10pt}

\textbf{Acknowledgements:} We thank Daniel B. Shaw and Jiarong Wu for helpful discussions regarding the numerical simulations. We thank Vatsal Sanjay for discussions regarding a modified Oldroyd-B implementation for Basilisk C.

\textbf{Funding:} This work was supported by the United States National Science Foundation grant 2242512 to L.D. and CBET-2246791 to H.A.S.


\appendix
\section{Grid refinement}\label{appAGrid}
\begin{figure}
\begin{center}
\includegraphics[width=0.8\textwidth]{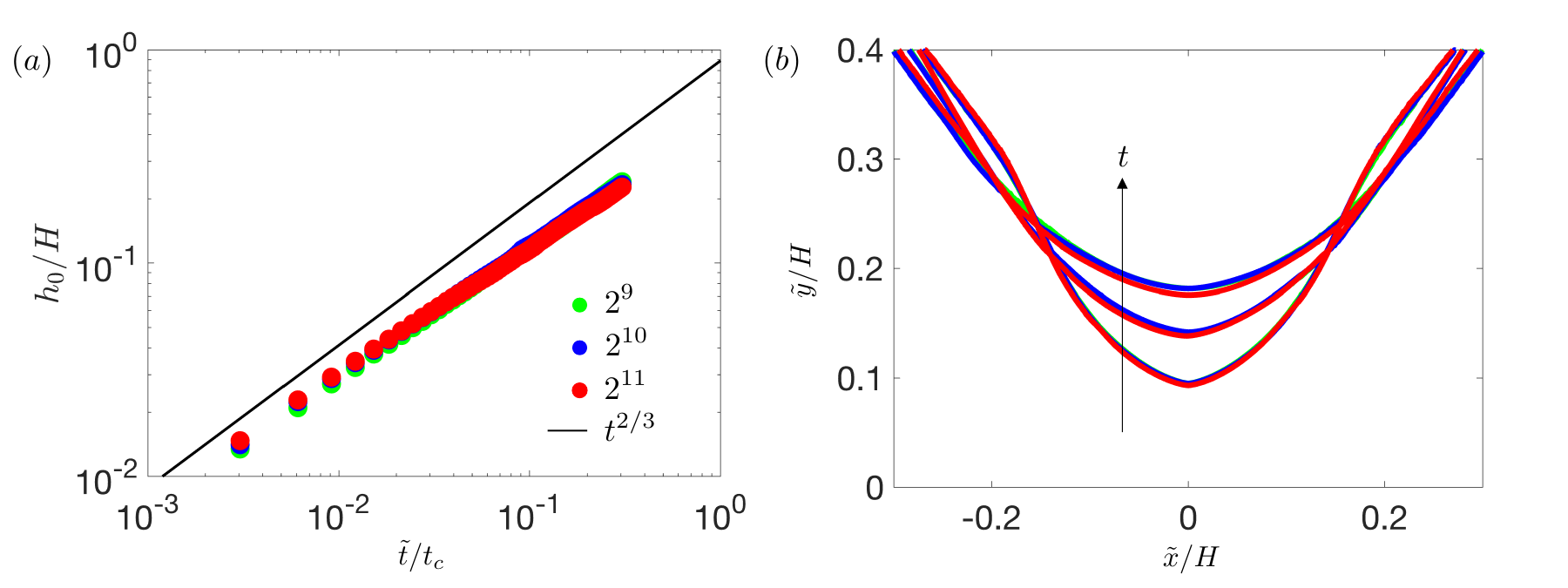}
\caption{\label{app_grid}  Results from varying the grid size in the simulations with $Oh = 0.09,~\theta=1.1,~De=0.01$, and $Ec=0.04$. (a) The temporal evolution of the height $h_0(t)$ at the coalescence point is shown for three different grid refinements: $2^{9},~2^{10}$, and $2^{11}$. The solid black like shows the $t^{2/3}$ scaling. (b) The shape of the interface for the three different grid refinements at three different time points. }
\end{center}
\end{figure}

We varied the grid refinement to ensure that the results are grid independent. Figure \ref{app_grid} shows the result from a grid independence study in which $Oh = 0.09,~\theta=1.1,~De=3.6$, and $Ec=0.04$. Figure \ref{app_grid}(a) shows the temporal evolution of the height $h_0(t)$ at the coalescence point for three different $L$ values.  Note that the lengths are rescaled by $H$ and the time is rescaled by the inertial-capillary time: $t_c = \sqrt{\rho H^{3}/\gamma}$. Figure \ref{app_grid}(b) shows the shape of the interface at different time steps for different $L$ values. The overlapping shape of the interface in Fig. \ref{app_grid} suggests that the data is well resolved at this grid refinement.

\section{Calculating average stress} \label{avestress-appB}
All the stress fields from the simulations were locally averaged over a square of size $\epsilon=0.007$ to make the results independent of the grid size. The average stress can be represented as

\begin{equation}
    \sigma = \frac{1}{\epsilon^2} \iint \hat{\sigma}~dS,
    \label{stress_average}
\end{equation}
where $dS$ is a square of length $\epsilon$. The value of $\epsilon$ was varied and was taken to be such that the result is independent of $\epsilon$ itself, and turned out to be slightly larger than the length of a $3 \times 3$ grid.

\section{Interfacial Curvature}
\label{curvature-appc}
\begin{figure}
\begin{center}
\includegraphics[width=0.8\textwidth]{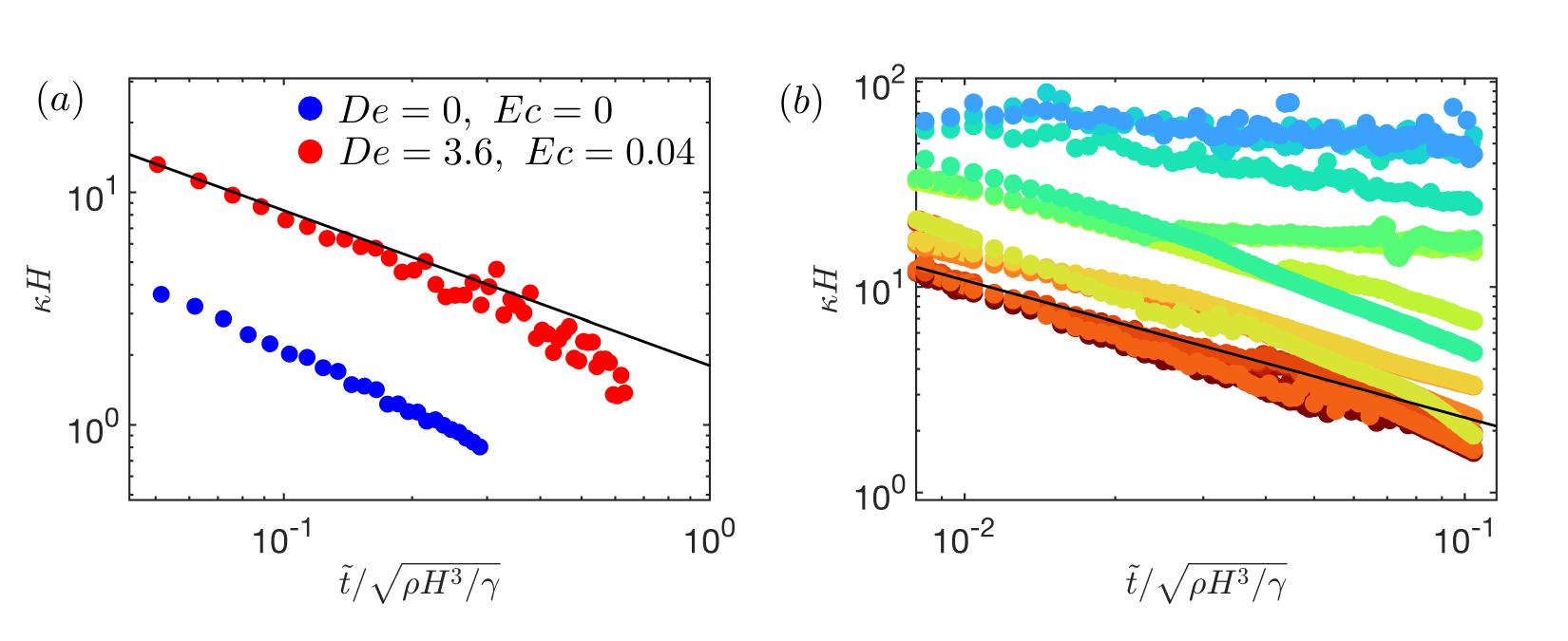}
\caption{\label{curvatures} The time evolution of the interfacial curvature near the coalescence point. (a) Curvature calculated from the experimental results shown in Fig. 6 in the main text, where the blue data corresponds to a Newtonian case with $Oh = 0.004,~\theta=1.1,~De=0$, and $Ec=0$ (experiment with water drops), and the red data corresponds to a non-Newtonian case with $Oh = 0.09,~\theta=1.1,~De=3.6$, and $Ec=0.04$ (experiment with 0.5\% PEO drops). The black line represents $t^{-2/3}$. (b) Curvature calculated from the simulation results. The data has the same legend as that shown in Fig. 7 in the main text. The black line represents $t^{-2/3}$.}
\end{center}
\end{figure}

The interfacial curvature in the $xz$-plane near the coalescence point ($x=0$) was calculated to quantify the sharpness of the interface. Given the shape of the interface $h(x,t)$, the curvature is defined as
\begin{equation}
    \kappa = \frac{h_{xx}}{(1+h_x^2)^{3/2}} \approx h_{xx},
    \label{stress_average}
\end{equation}
where the approximation is made since the slope vanishes at the coalescence point for symmetric coalescence. The curvature was calculated by fitting a parabola to the interface data, $h(x,t) = h_0(t) + \frac{1}{2}\kappa(t)~x^2$, considering the appropriate amount of data points that minimized the root-mean-square error of the fit. The results for the experiments are shown in Fig. \ref{curvatures}(a) and for the simulations are shown in Fig. \ref{curvatures}(b). Both the experimental and simulated data clearly show an increase in interfacial curvature in the presence of polymers. This quantification confirms the observation that was made in Fig. \ref{ch6fig1}(b)-(c).

The interfacial curvature for Newtonian drops scales as $\kappa \sim 1/h_0 \sim t^{-2/3}$, which is the black line in Fig. \ref{curvatures}. Note that as $De$ and $Ec$ increases, in Fig. \ref{curvatures}(b), the data deviates away from this scaling.

\section{FS-SS: Image processing}
\label{imageprocess-appD}
\begin{figure}
\begin{center}
\includegraphics[width=0.9\textwidth]{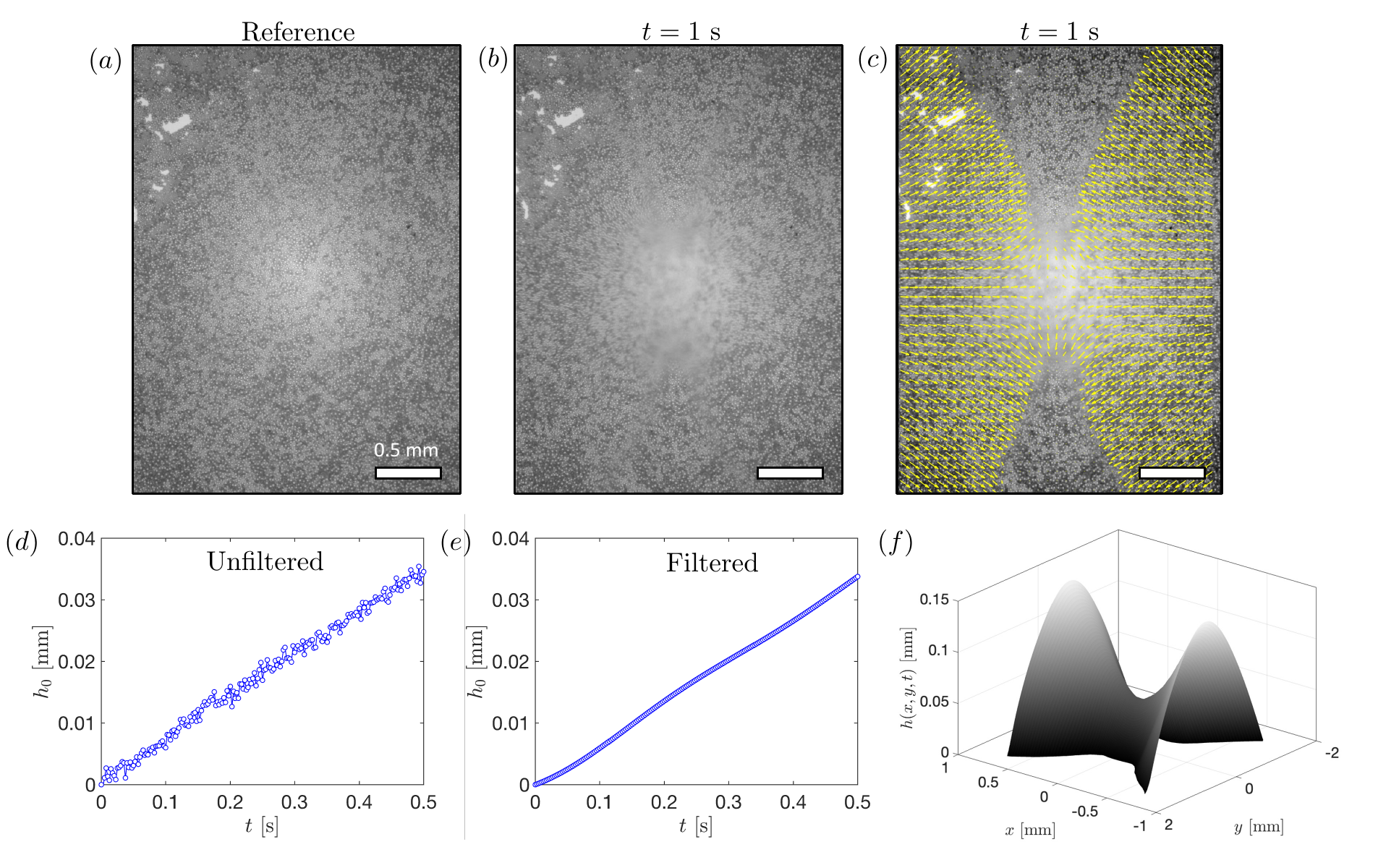}
\caption{\label{ch6_imageprocess} Image processing to derive height data from Free-Surface Synthetic Schlieren imaging. (a) Image showing the reference configuration of the dot pattern without a drop on top. (b) Snapshot of coalescence of 1\% PEO drops at $t=1$ s ($\theta \approx 11.7^{\circ}$). Notice that the image is slightly distorted compared to the reference image. (c) Displacement field from PIVLab for the $t=1$ s image showing the extend of the distortion caused by the drop on top of the dot pattern. All scale bars represent 0.5 mm. (d) Unfiltered and (e) filtered data showing the height evolution at the coalescence point. (e) Reconstructed 3D profile of the interface for $t=1$ s.  }
\end{center}
\end{figure}

Free-Surface Synthetic Schlieren (FS-SS) utilizes the fact that the curved shape of the drop acts as a lens to distort the image of the dot pattern that is placed under the drop. By measuring the distortion, or the displacement field of the dot pattern compared to a reference image, we can calculate the shape of the interface that caused the distortion. The displacement field is calculated using PIVlab \citep{thielicke2014pivlab}. Figure \ref{ch6_imageprocess}(a) shows a reference image of the dot pattern above which the coalescence experiment is performed. The dot pattern is produced by evaporating a dense suspension of poly-dispersed polystyrene particles. A microscope slide of thickness 1 mm is placed above the dot pattern, with approximately 0.3 mm space in between. Thus, the distance between the dot pattern and the plane of coalescence is $\Delta h = 1.3$ mm. The value of $\Delta h$ is used in image processing to calculate the interface profile. Figure \ref{ch6_imageprocess}(b) shows the distorted image of the dot pattern at $t=1$ s into the coalescence of 1\% PEO drops, and Fig. \ref{ch6_imageprocess}(c) shows the displacement field vectors for this frame. Note that all the calculated vectors are not shown in the image to avoid crowding.  

The displacement field is converted to the height profile of the interface using functions in the PIVMat toolbox in MATLAB \citep{pivmat}, which is based on the surface height reconstruction method delineated in Ref. \citep{moisy2009synthetic}. Figure \ref{ch6_imageprocess}(d) shows the height $h_0(t)$ at the coalescence point from the reconstructed interface profile data. The spatial resolution of our FS-SS experiments is of order $0.1~\mu$m and is a function of the imaging window size and the resolution of the camera. However, note that the data shows high-frequency oscillations. This noise is due to the internal fan of the high-speed camera which produces vibrations that are large enough to be observed within the spatiotemporal resolution that we operate in. We use a low-pass filter to attenuate the high-frequency oscillations, and the filtered data is shown in Fig. \ref{ch6_imageprocess}(e). No further smoothing is done on the experimental data. Figure \ref{ch6_imageprocess}(f) shows the reconstructed 3D shape of the interface corresponding to $t=1$ s.

\newpage
\bibliography{main}

@article{moisy2009synthetic,
  title={A synthetic Schlieren method for the measurement of the topography of a liquid interface},
  author={Moisy, Fr{\'e}d{\'e}ric and Rabaud, Marc and Salsac, K{\'e}vin},
  journal={Experiments in Fluids},
  volume={46},
  number={6},
  pages={1021--1036},
  year={2009},
  publisher={Springer}
}

@article{thielicke2014pivlab,
  title={PIVlab--towards user-friendly, affordable and accurate digital particle image velocimetry in MATLAB},
  author={Thielicke, William and Stamhuis, Eize},
  journal={Journal of Open Research Software},
  volume={2},
  number={1},
  year={2014},
  publisher={Ubiquity Press}
}

@article{ristenpart2006coalescence,
  title={Coalescence of spreading droplets on a wettable substrate},
  author={Ristenpart, WD and McCalla, PM and Roy, RV and Stone, HA},
  journal={Physical Review Letters},
  volume={97},
  number={6},
  pages={064501},
  year={2006},
  publisher={APS}
}

@article{hernandez2012symmetric,
  title={Symmetric and asymmetric coalescence of drops on a substrate},
  author={Hern{\'a}ndez-S{\'a}nchez, JF and Lubbers, LA and Eddi, Antonin and Snoeijer, JH},
  journal={Physical Review Letters},
  volume={109},
  number={18},
  pages={184502},
  year={2012},
  publisher={APS}
}

@article{eddi2013influence,
  title={Influence of droplet geometry on the coalescence of low viscosity drops},
  author={Eddi, A and Winkels, KG and Snoeijer, JH},
  journal={Physical Review Letters},
  volume={111},
  number={14},
  pages={144502},
  year={2013},
  publisher={APS}
}

@article{mitchinson2010saliva,
  title={Saliva at a stretch},
  author={Mitchinson, Andrew},
  journal={Nature},
  volume={465},
  number={7299},
  pages={701--701},
  year={2010},
  publisher={Nature Publishing Group UK London}
}

@article{datta2022perspectives,
  title={Perspectives on viscoelastic flow instabilities and elastic turbulence},
  author={Datta, Sujit S and Ardekani, Arezoo M and Arratia, Paulo E and Beris, Antony N and Bischofberger, Irmgard and McKinley, Gareth H and Eggers, Jens G and L{\'o}pez-Aguilar, J Esteban and Fielding, Suzanne M and Frishman, Anna and others},
  journal={Physical Review Fluids},
  volume={7},
  number={8},
  pages={080701},
  year={2022},
  publisher={APS}
}

@article{snoeijer2020relationship,
  title={The relationship between viscoelasticity and elasticity},
  author={Snoeijer, JH and Pandey, Anupam and Herrada, MA and Eggers, Jens},
  journal={Proceedings of the Royal Society A},
  volume={476},
  number={2243},
  pages={20200419},
  year={2020},
  publisher={The Royal Society Publishing}
}

@article{varma2020universality,
  title={Universality in coalescence of polymeric fluids},
  author={Varma, Sarath Chandra and Saha, Aniruddha and Mukherjee, Siddhartha and Bandopadhyay, Aditya and Kumar, Aloke and Chakraborty, Suman},
  journal={Soft Matter},
  volume={16},
  number={48},
  pages={10921--10927},
  year={2020},
  publisher={Royal Society of Chemistry}
}

@article{varma2021coalescence,
  title={Coalescence of polymeric sessile drops on a partially wettable substrate},
  author={Varma, Sarath Chandra and Saha, Aniruddha and Kumar, Aloke},
  journal={Physics of Fluids},
  volume={33},
  number={12},
  year={2021},
  publisher={AIP Publishing}
}

@article{dekker2022elasticity,
  title={When elasticity affects drop coalescence},
  author={Dekker, Pim J and Hack, Michiel A and Tewes, Walter and Datt, Charu and Bouillant, Ambre and Snoeijer, Jacco H},
  journal={Physical Review Letters},
  volume={128},
  number={2},
  pages={028004},
  year={2022},
  publisher={APS}
}

@article{chen2022probing,
  title={Probing the coalescence of non-Newtonian droplets on a substrate},
  author={Chen, Hao and Pan, Xiaolong and Nie, Qichun and Ma, Qianli and Fang, Haisheng and Yin, Zhouping},
  journal={Physics of Fluids},
  volume={34},
  number={3},
  year={2022},
  publisher={AIP Publishing}
}

@article{varma2022rheocoalescence,
  title={Rheocoalescence: Relaxation time through coalescence of droplets},
  author={Varma, Sarath Chandra and Rajput, Abhineet Singh and Kumar, Aloke},
  journal={Macromolecules},
  volume={55},
  number={14},
  pages={6031--6039},
  year={2022},
  publisher={ACS Publications}
}

@article{sivasankar2023coalescence,
  title={Coalescence of 3D Polymeric Drops in the Presence of In Situ Photopolymerization},
  author={Sivasankar, Vishal Sankar and Etha, Sai Ankit and Hines, Daniel R and Das, Siddhartha},
  journal={Macromolecules},
  year={2023},
  publisher={ACS Publications}
}

@inproceedings{Toms1948SomeOO,
  title={Some Observations on the Flow of Linear Polymer Solutions Through Straight Tubes at Large Reynolds Numbers},
  author={Benjamin A. Toms},
  year={1948},
  url={https://api.semanticscholar.org/CorpusID:139141851}
}

@article{berman1978drag,
  title={Drag reduction by polymers},
  author={Berman, Neil S},
  journal={Annual Review of Fluid Mechanics},
  volume={10},
  number={1},
  pages={47--64},
  year={1978},
  publisher={Annual Reviews 4139 El Camino Way, PO Box 10139, Palo Alto, CA 94303-0139, USA}
}

@article{anna2001elasto,
  title={Elasto-capillary thinning and breakup of model elastic liquids},
  author={Anna, Shelley L and McKinley, Gareth H},
  journal={Journal of Rheology},
  volume={45},
  number={1},
  pages={115--138},
  year={2001},
  publisher={The Society of Rheology}
}

@article{kaneelil2022three,
  title={Three-dimensional self-similarity of coalescing viscous drops in the thin-film regime},
  author={Kaneelil, Paul R and Pahlavan, Amir A and Xue, Nan and Stone, Howard A},
  journal={Physical Review Letters},
  volume={129},
  number={14},
  pages={144501},
  year={2022},
  publisher={APS}
}

@article{popinet2015basiliskquadtree,
  title={A quadtree-adaptive multigrid solver for the Serre--Green--Naghdi equations},
  author={Popinet, St{\'e}phane},
  journal={Journal of Computational Physics},
  volume={302},
  pages={336--358},
  year={2015},
  publisher={Elsevier}
}

@article{fattal2004constitutive,
  title={Constitutive laws for the matrix-logarithm of the conformation tensor},
  author={Fattal, Raanan and Kupferman, Raz},
  journal={Journal of Non-Newtonian Fluid Mechanics},
  volume={123},
  number={2-3},
  pages={281--285},
  year={2004},
  publisher={Elsevier}
}

@article{hao2007simulation,
  title={Simulation for high Weissenberg number: viscoelastic flow by a finite element method},
  author={Hao, Jian and Pan, Tsorng-Whay},
  journal={Applied Mathematics Letters},
  volume={20},
  number={9},
  pages={988--993},
  year={2007},
  publisher={Elsevier}
}

@article{popinet2009surftension,
  title={An accurate adaptive solver for surface-tension-driven interfacial flows},
  author={Popinet, St{\'e}phane},
  journal={Journal of Computational Physics},
  volume={228},
  number={16},
  pages={5838--5866},
  year={2009},
  publisher={Elsevier}
}

@article{stone2023note,
author ={Stone, Howard A. and Shelley, Michael J. and Boyko, Evgeniy},
title  ={A note about convected time derivatives for flows of complex fluids},
journal  ={Soft Matter},
year  ={2023},
month = {Jul},
volume  ={19},
issue  ={28},
pages  ={5353-5359},
publisher  ={The Royal Society of Chemistry},
}

@article{eggers1997ReviewModPhys,
  title={Nonlinear dynamics and breakup of free-surface flows},
  author={Eggers, Jens},
  journal={Reviews of Modern Physics},
  volume={69},
  number={3},
  pages={865},
  year={1997},
  publisher={APS}
}

@article{eddi2008wave,
  title={Wave propelled ratchets and drifting rafts},
  author={Eddi, Antonin and Terwagne, Denis and Fort, Emmanuel and Couder, Yves},
  journal={Europhysics Letters},
  volume={82},
  number={4},
  pages={44001},
  year={2008},
  publisher={IOP Publishing}
}

@article{datt2022thin,
  title={A thin-film equation for a viscoelastic fluid, and its application to the Landau--Levich problem},
  author={Datt, Charu and Kansal, Minkush and Snoeijer, Jacco H},
  journal={Journal of Non-Newtonian Fluid Mechanics},
  volume={305},
  pages={104816},
  year={2022},
  publisher={Elsevier}
}

@article{lee2002study,
  title={A study of viscoelastic free surface flows by the finite element method: Hele--Shaw and slot coating flows},
  author={Lee, Alex G and Shaqfeh, Eric SG and Khomami, Bamin},
  journal={Journal of Non-Newtonian Fluid Mechanics},
  volume={108},
  number={1-3},
  pages={327--362},
  year={2002},
  publisher={Elsevier}
}

@misc{pivmat,
  title = {PIVMat toolbox for MATLAB},
  howpublished = {\url{http://www.fast.u-psud.fr/pivmat}},
}

@article{turkoz2018axisymmetric,
  title={Axisymmetric simulation of viscoelastic filament thinning with the Oldroyd-B model},
  author={Turkoz, Emre and Lopez-Herrera, Jose M and Eggers, Jens and Arnold, Craig B and Deike, Luc},
  journal={Journal of Fluid Mechanics},
  volume={851},
  pages={R2},
  year={2018},
  publisher={Cambridge University Press}
}

@article{lopez2019adaptive,
  title={An adaptive solver for viscoelastic incompressible two-phase problems applied to the study of the splashing of weakly viscoelastic droplets},
  author={L{\'o}pez-Herrera, Jos{\'e}-Mar{\'\i}a and Popinet, St{\'e}phane and Castrej{\'o}n-Pita, Alfonso-Arturo},
  journal={Journal of Non-Newtonian Fluid Mechanics},
  volume={264},
  pages={144--158},
  year={2019},
  publisher={Elsevier}
}

@article{dixit2024viscoelastic,
  title={Viscoelastic Worthington jets$\backslash$\& droplets produced by bursting bubbles},
  author={Dixit, Ayush K and Oratis, Alexandros and Zinelis, Konstantinos and Lohse, Detlef and Sanjay, Vatsal},
  journal={arXiv preprint arXiv:2408.05089},
  year={2024}
}

@article{keller2000merging,
  title={Merging and wetting driven by surface tension},
  author={Keller, Joseph B and Milewski, Paul A and Vanden-Broeck, Jean-Marc},
  journal={European Journal of Mechanics-B/Fluids},
  volume={19},
  number={4},
  pages={491--502},
  year={2000},
  publisher={Elsevier}
}

@article{eggers2024reviewcoalescence,
  title={Coalescence dynamics},
  author={Eggers, Jens and Sprittles, James E and Snoeijer, Jacco H},
  journal={Annual Review of Fluid Mechanics},
  volume={57},
  year={2024},
  publisher={Annual Reviews}
}

@article{varma2022elasticity,
  title={Elasticity can affect droplet coalescence},
  author={Varma, Sarath Chandra and Dasgupta, Debayan and Kumar, Aloke},
  journal={Physics of Fluids},
  volume={34},
  number={9},
  year={2022},
  publisher={AIP Publishing}
}

@article{mostert2022high,
  title={High-resolution direct simulation of deep water breaking waves: transition to turbulence, bubbles and droplets production},
  author={Mostert, Wouter and Popinet, St{\'e}phane and Deike, Luc},
  journal={Journal of Fluid Mechanics},
  volume={942},
  pages={A27},
  year={2022},
  publisher={Cambridge University Press}
}

@article{van2022fourth,
  title={A fourth-order accurate adaptive solver for incompressible flow problems},
  author={van Hooft, J Antoon and Popinet, St{\'e}phane},
  journal={Journal of Computational Physics},
  volume={462},
  pages={111251},
  year={2022},
  publisher={Elsevier}
}

@article{pilz2007critical,
  title={On the critical bubble volume at the rise velocity jump discontinuity in viscoelastic liquids},
  author={Pilz, Christian and Brenn, G{\"u}nter},
  journal={Journal of non-newtonian fluid mechanics},
  volume={145},
  number={2-3},
  pages={124--138},
  year={2007},
  publisher={Elsevier}
}

@article{astarita1965motion,
  title={Motion of gas bubbles in non-Newtonian liquids},
  author={Astarita, Gianni and Apuzzo, Gennaro},
  journal={AIChE Journal},
  volume={11},
  number={5},
  pages={815--820},
  year={1965},
  publisher={Wiley Online Library}
}

@article{fraggedakis2016velocity,
  title={On the velocity discontinuity at a critical volume of a bubble rising in a viscoelastic fluid},
  author={Fraggedakis, D and Pavlidis, M and Dimakopoulos, Y and Tsamopoulos, J},
  journal={Journal of Fluid Mechanics},
  volume={789},
  pages={310--346},
  year={2016},
  publisher={Cambridge University Press}
}

@article{harlen1990high,
  title={High-Deborah-number flows of dilute polymer solutions},
  author={Harlen, OG and Rallison, JM and Chilcott, MD},
  journal={Journal of non-newtonian fluid mechanics},
  volume={34},
  number={3},
  pages={319--349},
  year={1990},
  publisher={Elsevier}
}

@article{cressely1979lines,
  title={Localized flow birfringence lines in a device with two counter-rotating rollers},
  author={Cressely, R and Hocquart, H and Scrivener, O},
  journal={Optica Acta: International Journal of Optics},
  volume={26},
  number={9},
  pages={1173--1181},
  year={1979},
  publisher={Taylor \& Francis}
}

@article{farrell1978observation,
  title={The observation of high polymer chain extension with two counter-rotating rollers},
  author={Farrell, CJ and Keller, A},
  journal={Colloid and Polymer Science},
  volume={256},
  pages={966--969},
  year={1978},
  publisher={Springer}
}

@article{rostami2025coalescence,
  title={Coalescence of viscoelastic drops on a solid substrate},
  author={Rostami, Peyman and Erb, Alexander and Azizmalayeri, Reza and Steinmann, Johanna and Stark, Robert W and Auernhammer, G{\"u}nter K},
  journal={Physical Review Fluids},
  volume={10},
  number={6},
  pages={063603},
  year={2025},
  publisher={APS}
}

@article{tirtaatmadja2006drop,
  title={Drop formation and breakup of low viscosity elastic fluids: Effects of molecular weight and concentration},
  author={Tirtaatmadja, Viyada and McKinley, Gareth H and Cooper-White, Justin J},
  journal={Physics of fluids},
  volume={18},
  number={4},
  year={2006},
  publisher={AIP Publishing}
}

\end{document}